\documentclass[usenatbib, useAMS]{mn2e}
\usepackage{color}
\usepackage{multirow}
\usepackage{graphicx}
\usepackage{amsmath}
\usepackage{natbib}
\usepackage{myaasmacros}
\usepackage{enumerate}
\usepackage{bm}
\usepackage[english]{babel}
\usepackage{mathptmx}
\usepackage{tabularx}
\newcommand{\Msol}{\mathrm{M}_{\odot}}
\newcommand{\kpc}{\mathrm{kpc}}

\newcommand{\s}{\mathrm{s}}
\newcommand{\km}{\mathrm{km}}
\newcommand{\kms}{\mathrm{km\,s^{-1}}}

\newcommand{\jcirc}{j_{\mathrm{circ}}}
\newcommand{\pc}{\mathrm{pc}}

\newcommand{\Myr}{\mathrm{Myr}}
\newcommand{\dd}{\mathrm{d}}
\newcommand{\Gyr}{\mathrm{Gyr}}

\renewcommand{\vec}[1]{\bm{#1}}
\newcommand{\tdyn}{t_{\mathrm{dyn}}}
\newcommand{\vmax}{v_{\mathrm{max}}}
\newcommand{\rmax}{r_{\mathrm{max}}}

\def\lsim{\mathrel{\rlap{\lower3pt\hbox{$\sim$}}
    \raise1pt\hbox{$<$}}}                
\def\gsim{\mathrel{\rlap{\lower3pt\hbox{$\sim$}}
    \raise1pt\hbox{$>$}}}                

\begin{document}
\pubyear{2015} \title [Milking the spherical cow]{Milking the
  spherical cow --- \newline on aspherical dynamics in spherical coordinates}
\author[A. Pontzen et al]{Andrew Pontzen$^{1,2,3,4}$,
  Justin~I.~Read$^5$, Romain Teyssier$^{6}$, Fabio Governato$^{7}$,
  \and Alessia Gualandris$^{5}$, Nina Roth$^{1}$,
  Julien~Devriendt$^{2}$
  \\
  $^1$ {Department of Physics and Astronomy, University College
    London,
    London WC1E 6BT} \\
  $^2$ {Oxford Astrophysics, Denys Wilkinson Building, Keble Road,
    Oxford, OX1 3RH} \\
  $^3$ {Balliol College, Broad Street, Oxford, OX1 3BJ} \\
  $^4$ {Email: a.pontzen@ucl.ac.uk} \\
  $^5$ {Department of Physics, University of Surrey, Guildford, GU2
    7XH, Surrey, UK} \\
  $^6$ {Institute for Theoretical Physics, University of Zurich,
    CH-8057
    Z\"{u}rich, Switzerland} \\
  $^7$ {Astronomy Department, University of Washington, Seattle, WA
    98195, US} \\
}

\date{ Received ---; published---. }
\maketitle

\vspace{-1cm}
\begin{abstract}
  Galaxies and the dark matter halos that host them are not
  spherically symmetric, yet spherical symmetry is a helpful
  simplifying approximation for idealised calculations and
  analysis of observational data. The assumption leads to an exact
  conservation of angular momentum for every particle, making the
  dynamics unrealistic. But how much does that inaccuracy matter in
  practice for analyses of stellar distribution functions,
  collisionless relaxation, or dark matter core-creation?

  We provide a general answer to this question for a wide class of
  aspherical systems; specifically, we consider distribution functions
  that are ``maximally stable'', {\it i.e.} that do not evolve at
  first order when external potentials (which arise from baryons,
  large scale tidal fields or infalling substructure) are applied. We
  show that a spherically-symmetric analysis of such systems gives
  rise to the false conclusion that the density of particles in phase
  space is ergodic (a function of energy alone).

  Using this idea we are able to demonstrate that: (a) observational
  analyses that falsely assume spherical symmetry are made more
  accurate by imposing a strong prior preference for near-isotropic
  velocity dispersions in the centre of spheroids; (b) numerical
  simulations that use an idealised spherically-symmetric setup can
  yield misleading results and should be avoided where possible; and
  (c) triaxial dark matter halos (formed in collisionless cosmological
  simulations) nearly attain our maximally-stable limit, but their
  evolution freezes out before reaching it.  \vspace{0.5cm}
\end{abstract}

\section{Introduction}

Spherical symmetry is a foundational assumption of many dynamical
analyses. The primary motivation is simplicity, since few astronomical
objects are actually spherical. For example, observations and
simulations both suggest that gravitational potential wells generated
by dark matter halos are typically triaxial
\cite[e.g.][]{1991ApJ...378..496D,1996MNRAS.281..716C,2002ApJ...574..538J,2005ApJ...629..781K,2007MNRAS.377...50H,2012JCAP...05..030S,Loebman12triaxialMW}.
Characterising dark matter halos by spherically-averaged densities and
velocities
\citep[e.g.][]{1991ApJ...378..496D,1996ApJ...462..563N,TaylorNavarro01NFWPhaseSpace,2009MNRAS.398L..21S}
at best tells only part of the story. At worst, it could be severely
misleading.

The question of whether baryonic processes can convert dark matter
cusps into cores \citep{pontzen2014nature} provides one motivation for
a detailed study of the relationship between spherical and
near-spherical dynamics. To explain why, we need to look forward to
some of our results. Later in this paper, we cut a dark matter halo
out of a cosmological simulation, then match it to an exactly
spherical halo with an identical density and velocity anisotropy
profile. This gives us two easy-to-compare equilibrium structures --
the first triaxial, the second spherical -- to perform a dynamical
comparison.  We expose each to the same time-varying gravitational potential,
mimicking the effects of stellar feedback (there are no actual baryons
in these runs). After $1\,\Gyr$, the triaxial halo's averaged density
profile flattens into a convincing dark matter core, but the spherical
halo maintains its cusp (see Figure 1).

This example, which is fully explored in Section \ref{sec:cusp-core},
illustrates how it is dangerous to use spherically-symmetric
simulations to infer anything about dynamics --- even
spherically-averaged dynamics --- in the real universe. A spherical
system does not evolve in the same way as the spherical averages of a
triaxial system.

Ignoring asphericity can also lead to observational biases
\cite[e.g.][]{HayashiNavarro06,Corless07triaxlens}. From a dynamical
standpoint, the nature of orbits in triaxial potentials is
fundamentally different from those in spherical potentials: although
the total angular momentum of any self-gravitating system must always
be conserved, it is only in the spherical case that this conservation
holds for individual particles. Conversely, a large fraction of dark
matter particles near the centre of cosmological halos will be on box
orbits which do not conserve their individual angular momenta
\citep{1983ApJ...267..571D,1996ApJ...471...82M,2001ApJ...549..862H,2007ApJ...670.1027A}. One
practical consequence is that asphericity may be responsible for filling the
loss cones of supermassive black holes at the centre of the
corresponding galaxies \citep{2004MerritLossCone}.

Finally, it is known that asphericity plays a fundamental role in
setting the equilibrium density profile during gravitational cold
collapse \cite[e.g.][]{1999ApJ...517...64H}. The underlying process is
known as the radial orbit instability or ROI
\citep{1973A&A....24..229H,1991MNRAS.248..494S,1999ApJ...517...64H,2006ApJ...653...43M,2008ApJ...685..739B,2009ApJ...704..372B,Marechal09ROI};
a related effect was discussed by \cite{2007ApJ...670.1027A}. The name
arises because particles on radial orbits are perturbed onto more
circular trajectories. At the same time, the density distribution
becomes triaxial. Even in the case of a uniform spherical collapse,
this symmetry-breaking process is still triggered, presumably by
numerical noise; the tangential component of forces must be
unphysically suppressed for the system to remain spherical
\citep{1999ApJ...517...64H,2006ApJ...653...43M}.

Despite all this, assuming spherical symmetry is very tempting because
it makes life so much easier. Defining spherically-averaged quantities
is a well-defined and sensible procedure even if we actually have the
full distribution function in hand (as in simulations): departures
from spherical symmetry are sufficiently small that different
averaging procedures lead to consistent results
\citep{saharead09}. Additionally, when an aspherical halo is in
equilibrium, we have shown numerically that a ``sphericalised''
version of it is also in equilibrium (see Appendix B of
\citeauthor{2013MNRAS.430..121P} 2013). This is helpful because it
allows one to make a meaningful analysis in spherical coordinates,
even when the system is aspherical. But it breaks down when
out-of-equilibrium processes are included, as in the
stellar-feedback-driven core-creation example above.

\begin{figure}
\includegraphics[width=0.49\textwidth]{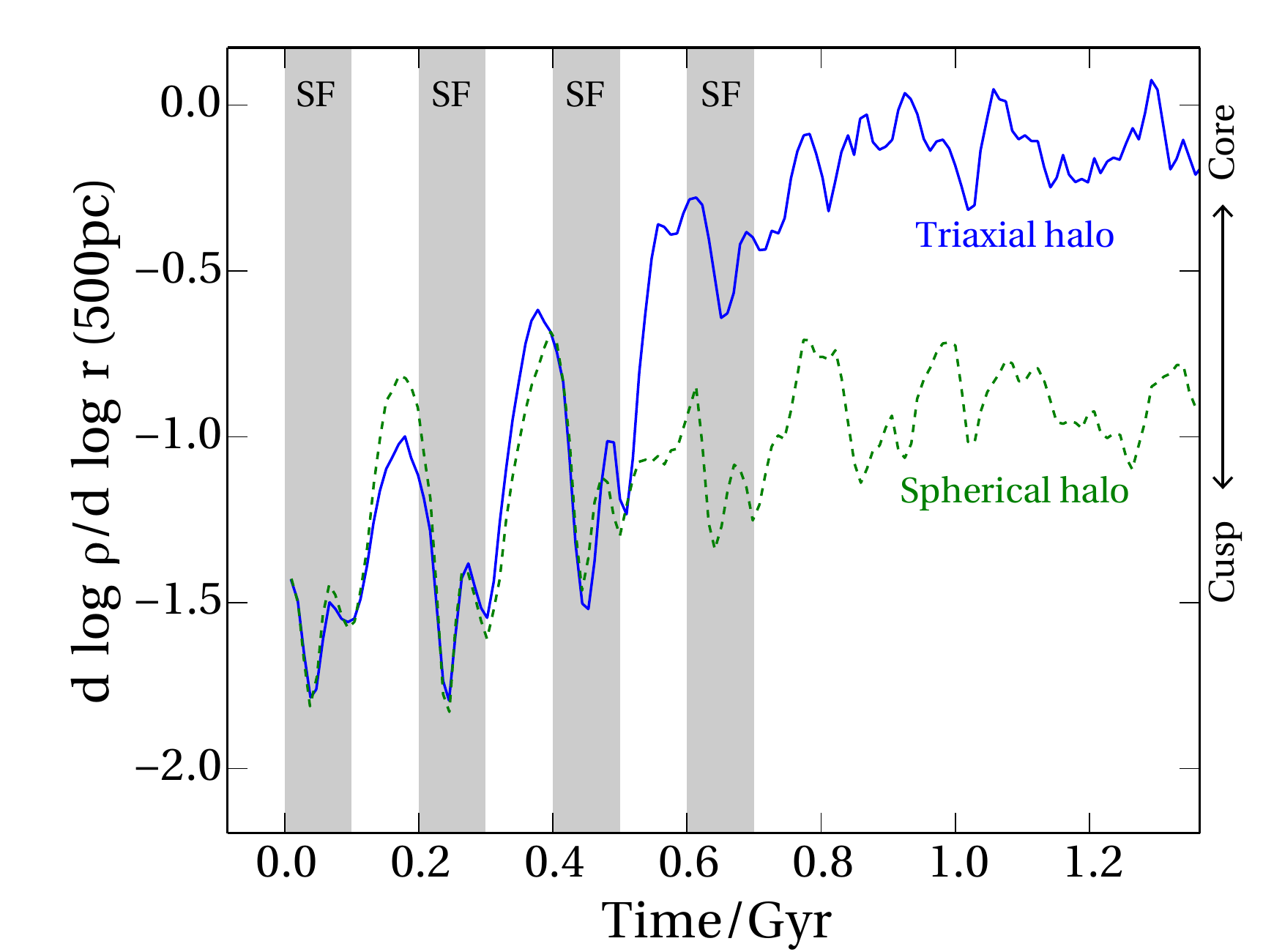}
\caption{One motivation for studying the relationship between
  spherical and aspherical dynamics is that the conversion of a dark
  matter cusp into a core by baryonic processes is qualitatively
  different in the two cases. Here we show the inner log density slope
  from a numerical experiment on two halos. One is spherical (dashed
  line) and the other triaxial (solid line) but their
  spherically-averaged properties are initially identical.  An external
  potential has been added at the centre during the times indicated by
  the grey bands, with the fluctuations mimicking stellar
  feedback. The triaxial halo develops a clear core, whereas the
  spherical halo almost maintains its central density cusp. A complete
  description and analysis is given in Section~\ref{sec:cusp-core}. }
\end{figure}

The present paper formalizes the idea of spherical analysis performed
on aspherical systems and follows through the consequences. We will
study equilibrium distribution functions in nearly, but not exactly,
spherically-symmetric potentials, and focus on maximally-stable
systems (which we define as being stable against all possible external
linear perturbations).  We will find that, in spherical coordinates,
such systems appear to be ``ergodic" (meaning that their distribution
functions depend on energy alone) because the individual
particles move randomly in angular momentum while maintaining a
near-constant energy. It is important to emphasise that this describes
the appearance of the system when analysed in a spherical coordinate
system and the true system need not be chaotic for the result to hold,
provided any isolating integrals are not closely related to angular
momentum.

The formal statement of this idea is derived in
Section~\ref{sec:most-stable-system}. A brief overview of the required
background is given in Appendix~\ref{sec:background}, and a
second-order derivation of the evolution is given in
Appendix~\ref{sec:f0-evolution}. Section~\ref{sec:ROI-connection}
outlines the practical consequences, starting by recasting and
extending the phenomenology of the radial orbit instability. We
describe an immediate implication for observational studies of
aspherical systems, namely a new way to break the anisotropy
degeneracy. Finally we return to the motivating problem above and
explain why triaxial systems can undergo cusp-core transitions more
easily than spherical systems.  Section~\ref{sec:conclusions}
concludes and points to open questions and future work.


\section{Aspherical dynamics in spherical coordinates}\label{sec:most-stable-system}

In this section we consider an aspherical system which is maximally
stable against external linear perturbations. We assume that an
observer of this system analyses it assuming spherical symmetry. We
will show that this observer (falsely) concludes that the system is
ergodic, {\it i.e.} that the density of particles in phase space is a
function of energy alone. The derivation requires the use of
action-angle coordinates; a crash course is provided in Appendix~\ref{sec:background}.

\subsection{Single particles}

Given any near-spherical system, the Hamiltonian in the spherical
action-angle variables is
\begin{equation}
H(\vec{J}, \vec{\Theta}) = H_0(\vec{J}) + \delta H(\vec{J}, \vec{\Theta})\textrm{,}\label{eq:H-perturbed}
\end{equation}
where $H_0$ is the sum of kinetic and potential energies in the
spherical background, $\vec{J}=(J_r,j,j_z)$ is the vector of spherical
actions (see Appendix \ref{sec:background}), $\vec{\Theta}$ is the
vector of spherical angles and $\delta H$ contains the perturbation
(which includes the aspherical correction to the potential). The
orbit of a particle in exact spherical symmetry, $\delta H=0$, is
described by Hamilton's equations:
\begin{equation}
\dot{\vec{J}}_0 = -\frac{\partial H_0}{\partial \vec{\Theta}} = 0 \textrm{;} \hspace{1cm}
\dot{\vec{\Theta}}_0 = \left. \frac{\partial H_0}{\partial \vec{J}} \right|_{\vec{J}=\vec{J}_0} \equiv
\Omega_0(\vec{J}_0)\textrm{,}
\end{equation}
which defines the constant background orbital frequencies
$\Omega_0(\vec{J}_0)$.  The expressions $\vec{J}_0$, $\vec{\Theta}_0$
and $\Omega_0(\vec{J}_0)$ will be used throughout to refer to the background
($\delta H=0$) solution.  This algebraically simple form of the
equations of motion is the reason for using action-angle variables,
since it immediately integrates to
\begin{equation}
\vec{J}_0(t) = \vec{J}_0 = \textrm{constant; }\hspace{0.5cm} \vec{\Theta}_0(t) =
\vec{\Theta}_0(0) + \vec{\Omega}_0(\vec{J}_0) t\textrm{,}\label{eq:background-solution}
\end{equation}
where $\vec{J}_0$ and $\vec{\Theta}_0(0)$ specify the initial
action and angle coordinates of the orbit.

We now consider the effect of the aspherical correction to the
potential encoded in $\delta H$, using standard Hamiltonian
perturbation theory
\cite[e.g.][]{1992rsm..book.....L,BinneyTremaine2008}. First, taking
advantage of the angle coordinates $\vec{\Theta}$ being periodic in
$2\pi$, $\delta H$ is expressed as
\begin{equation}
\delta H(\vec{J},\vec{\Theta}) = \sum_{\vec{n}} \delta H_{\vec{n}}(\vec{J}) e^{i\vec{n}\cdot\vec{\Theta}}\textrm{.}\label{eq:H-expansion}
\end{equation}
This equation states that, at any fixed $\vec{J}$, one can expand
the periodic $\Theta$ dependence in a Fourier series without loss of generality.

We are interested in the evolution of $\vec{J}$ at first order in
the perturbation. Hamilton's relevant equation now reads:
\begin{equation}
\dot{\vec{J}} = - \frac{\partial H}{\partial \vec{\Theta}} = - \sum_{\vec{n}} i \vec{n}
\delta H_{\vec{n}}(\vec{J}) e^{i \vec{n} \cdot \vec{\Theta}}\textrm{.}\label{eq:hamilton-deltaH}
\end{equation}
Because $\delta H$ is small, the result to first-order accuracy
is given by substituting the zero-order solution
\eqref{eq:background-solution} into equation
\eqref{eq:hamilton-deltaH} and integrating to give
\begin{equation}
\vec{J}(t) = \vec{J}_0 - \sum_{\vec{n}} \frac{\vec{n}}{\vec{n} \cdot \vec{\Omega}_0} \delta H_{\vec{n}}(\vec{J}_0) e^{i \vec{n} \cdot \vec{\Theta}_0(t)} + \cdots \textrm{.}\label{eq:J-first-order}
\end{equation}
Consequently as $\vec{n} \cdot \vec{\Omega}_0 \to 0$, the
linear-order correction to the orbit of a particle can become large
even if the aspherical correction to the potential ($\delta H$) is
small, an effect known as resonance
\citep{BinneyTremaine2008}. Consider now the evolution of the
background energy along the perturbed trajectory, given by
\begin{equation}
H_0(\vec{J}(t)) \simeq H_0(\vec{J}_0) + \frac{\partial H_0}{\partial \vec{J}} \cdot \left(\vec{J}(t) - \vec{J}_0\right) + \cdots\textrm{,}
\end{equation}
where we have Taylor-expanded to first order around $\vec{J}_0$. Substituting equation \ref{eq:J-first-order} for $J(t)$, there is a cancellation between numerator and denominator:
\begin{equation}
H_0(\vec{J}(t)) \simeq H_0(\vec{J}_0) - \sum_{\vec{n}}\delta H_{\vec{n}}(\vec{J}_0) e^{i \vec{n} \cdot \vec{\Theta}_0(t)} +\cdots\label{eq:H0-variation-small}
\end{equation}
and so the fractional variation in $H_0$ remains small, even if
$\vec{J}$ changes significantly over time.  In other words, according
to linear perturbation theory, particles migrate large distances in
$\vec{J}$ along surfaces of approximately constant background energy
$H_0$.  One can verify this constrained-migration prediction
in numerical simulations of dark matter halos -- an explicit
demonstration is given in Appendix \ref{sec:Jr-j-halos}.  This gives
us some intuition for the result to come: a population of particles
will seem to ``randomise" their actions (including angular momentum),
but not their energy distribution.

The extent of the migration will depend on the nature of the potential
in which a particle orbits. To quantify this requires going beyond
linear perturbation theory and is the subject of ``KAM theory'' after
\cite{kolmogorov1954}, \cite{arnold1963} and \cite{moser1962}; see
e.g. \cite{BinneyTremaine2008,goldstein2002classical,1992rsm..book.....L}
for introductions. Colloquially the result is that for any given small
perturbation the migration of typical orbits is also small. Arnold
diffusion offers the most famous route to more significant diffusion
through action space \cite[see][]{1992rsm..book.....L}; but in our
case, there is a more immediate reason why the KAM result does not in
fact hold. Specifically, KAM theory relies on the frequencies
$\vec{\Omega}(\vec{J})$ being non-degenerate -- i.e. that any change
in the action leads also to a change in the frequencies, thus shutting
off resonant migration.  In smooth potentials, $\Omega$ is
almost a function of energy alone (see Appendix \ref{sec:Jr-j-halos})
and so the migration can be substantial.

Overall we informally expect particles to redistribute themselves
randomly within the action shell of fixed background energy until they
are evenly spread, implying a distribution function that appears
ergodic in a spherical analysis. This does not imply the orbits are
chaotic in the traditional sense; it is only because we are analysing
an aspherical system in spherical coordinates that the phenomenon
arises. With this in mind, we now turn to a more formal demonstration
of the result.

\subsection{The distribution function}\label{sec:population}

So far we have discussed how a single particle orbiting in a mildly aspherical potential does not conserve its spherical actions (e.g. angular momentum). We informally suggested that a population will appear to `randomise' the spherical actions at fixed energy. We now show more formally that a distribution function of particles subject to aspherical perturbations will be most stable when it is spread evenly on surfaces of constant $H_0$.

We start by decomposing the true distribution function of particles in phase space,
$f$, in terms of a spherical background $f_0$ and a perturbation
$\delta f$. To make sure the split between spherical background and aspherical
perturbation is uniquely defined, we take $f_0$ as the distribution function obtained
when we perform a naive analysis averaging out the aspherical contribution:
\begin{equation}
f_0(\vec{J}) \equiv \frac{1}{(2\pi)^3} \int \dd^3\Theta f(\vec{J},\vec{\Theta})\textrm{.}\label{eq:f0_from_spherical_average}
\end{equation}
By Jean's Theorem, $f_0$ is an equilibrium distribution function in the spherical background because it is constructed from spherical invariants $\vec{J}$ alone.
Analogous to equations \eqref{eq:H-perturbed} and
\eqref{eq:H-expansion} one can write the full distribution function $f$ as
\begin{align}
f(\vec{J},\vec{\Theta}) & = f_0(\vec{J}) + \delta f(\vec{J}, \vec{\Theta}) \nonumber \\
& = f_0(\vec{J}) + \sum_{\vec{n}} \delta f_{\vec{n}}(\vec{J}) e^{i
  \vec{n} \cdot \vec{\Theta}}\textrm{.}
\end{align}

The whole $f$ is to be in
equilibrium in the true system, $\partial f/\partial t=0$. We can turn
this into an explicit condition on $f$ using the
the collisionless Boltzmann equation,
\begin{equation}
0=\frac{\partial f}{\partial t} = [H,f] \equiv \frac{\partial
  H}{\partial \vec{\Theta}} \cdot \frac{\partial f}{\partial \vec{J}}
- \frac{\partial
  H}{\partial \vec{J}} \cdot \frac{\partial f}{\partial \vec{\Theta}}\textrm{.}
\end{equation}
Expanding to linear order in $H$ and $f$ gives the condition
\begin{equation}
\sum_{\vec{n}} \left( \vec{\Omega}_0(\vec{J}) \cdot \vec{n}
\delta f_{\vec{n}} (\vec{J}) - \frac{\partial f_0}{\partial \vec{J}}
\cdot \vec{n} \,\delta H_{\vec{n}}(\vec{J}) \right) e^{i \vec{n} \cdot
\vec{\Theta}} = 0\label{eq:aspherical-evolution}\textrm{.}
\end{equation}
The different $\vec{\Theta}$ dependence of each term in the sum means
that the term in brackets must be zero for each $\vec{n}$.
In particular, for the resonant terms $\vec{n}_{\perp}$ where $\vec{\Omega}_0 \cdot \vec{n}_{\perp}=0$ one has the condition
\begin{equation}
\left(\frac{\partial f_0}{\partial \vec{J}}\cdot \vec{n}_{\perp}\right) \,\delta H_{\vec{n}_{\perp}}(\vec{J})=0\textrm{.}\label{eq:sphergodicity}
\end{equation}
A sufficient condition for stability of $\delta f$ is therefore that $f_0$ is a function only of $H_0$,
since then $\partial f_0 / \partial \vec{J} =  \vec{\Omega}_0\, \dd f_0 / \dd H_0$ and consequently the dot product in equation \eqref{eq:sphergodicity} vanishes.

This is the core result claimed at the start of the section: $f_0 = f_0(H_0)$, {\it i.e.} the distribution function implied by a spherical analysis appears to be ergodic.
It is not a necessary condition for achieving equilibrium, since for any given aspherical system certain $\delta H_{\vec{n}_{\perp}}$ will be zero. Rather, the result should be read as applying to the maximally stable distribution function -- a distribution function that does not evolve under any linear perturbation to its potential.

We again emphasise that the distribution function $f_0$ is
fictional. There is no sense in which the true distribution function,
$f$, is actually ergodic. The statement is about how the system
appears to be when it is analysed using spherically averaged
quantities, equation \eqref{eq:f0_from_spherical_average}. Yet, it
establishes a way in which we can understand these
spherically-averaged quantities in a systematic way -- the system is
most stable if it {\it appears} ergodic, regardless of what the
underlying dynamics is really up to. In the remainder of this paper we
will refer to such systems as `spherically ergodic'.

\subsection{Testable predictions}\label{sec:sphergodic-predictions}

We have established that systems which appear to be ergodic in a
spherical analysis are maximally stable. Now we need to devise a
connection to observable or numerically-measurable quantities.

A distribution function $f(H_0)$ that is truly a function of energy
alone has an isotropic velocity distribution
\citep{BinneyTremaine2008}.  To test for isotropy, one calculates
$\beta(r)$ according to the usual spherically-averaged definition
\begin{equation}
\beta(r) = 1 - \frac{\langle v_t^2\rangle(r)}{2 \langle v_r^2 \rangle(r)}
\end{equation}
where $v_r$ is a particle's radial velocity, $v_t$ its tangential
velocity and the angle-bracket averages are taken in radial
shells. For a population on radial orbits, $\beta(r)=1$;
conversely for purely circular motion, $\beta(r)=-\infty$. Between
these two extremes, an ergodic population has $\beta(r)=0$ \citep{BinneyTremaine2008}.

Intuitively, a spherically ergodic system (in the sense defined in the
previous section) should therefore be approximately isotropic. However
one has to handle that expectation with a little care because the true
population $f$ is {\it not} ergodic and the measured velocity
dispersions, even in spherical polar coordinates, may inherit
information from $f$ that is not present in $f_0$.

Instead we will now construct a more rigorously justifiable, slightly different statement that still connects spherically ergodic populations to velocity isotropy. Measuring the mean of any function of the spherical actions $q(\vec{J})$, we obtain
\begin{equation}
\int \dd^3 J\, \dd^3 \Theta\, f(\vec{J},\vec{\Theta}) q(\vec{J}) = \int \dd^3 J \, f_0\left(H_0(\vec{J})\right) q(\vec{J})\textrm{,}\label{eq:q-average}
\end{equation}
an exact result. Therefore any statement about averages over spherical actions  automatically knows only about $f_0$ -- the spherical part of the distribution function. This allows us to derive unambiguous implications of a spherically ergodic population.

The most familiar action is the specific scalar angular momentum
$j$. Because it is a scalar for each particle, averages over this
quantity do not express anything about a net spin of the halo but
rather about the mix of circular and radial orbits, just like the
traditional velocity anisotropy. Radially-biased populations have
$\langle j\rangle\simeq 0$ whereas populations on circular orbits have
$\langle j \rangle= j_c$, where $j_c$ is the maximum angular
momentum available at a given energy. So, velocity anisotropy can be
conveniently represented in terms of the mean scalar angular momentum.

We can go further and calculate a function, $\langle j\rangle(E)$, where
the average is taken only over particles at a particular specific energy.
This quantity
can be represented in terms of the ratio of two integrals of the form
\eqref{eq:q-average}:
\begin{align}
\langle j \rangle(E) &= \frac{\iiint \dd J_r\, \dd j\, \dd j_z\,
f_0(H_0) j \delta(H_0-E)}{\iiint \dd J_r\, \dd j\, \dd j_z\,
f_0(H_0) \delta(H_0-E)}\textrm{.}
\end{align}
The triple integral ranges over the physical phase space coordinates:
$0\le J_r < \infty$, $0 \le j < \infty$, $-j \le j_z \le j $. One can
immediately perform the $j_z$ integrals; then the $J_r$ integral can
be completed by changing variables to $H_0$ (recalling $\partial
H_0/\partial J_r \equiv \Omega_r$) and consuming the $\delta$
function. After this manipulation $j$ can only range between $0$ and
$j_c(E)$ where $j_c(E)$ is the specific angular momentum corresponding
to a circular orbit with specific energy $E$; there are no physical orbits with more angular momentum at the specified $E$. The final, exact result is:
\begin{align}
\langle j \rangle(E) &= \left.\int_0^{j_c(E)} \dd j\, \Omega_r(E,j)^{-1}\, j^2\right/\int_0^{j_c(E)} \dd j\, \Omega_r(E,j)^{-1}\, j\textrm{.} \label{eq:mean-j-expression}
\end{align}
We now have a firm prediction for spherically ergodic
populations. Namely, if we bin particles in $E$ and measure $\langle j
\rangle(E)$ in each bin, the results should be predicted by equation
\ref{eq:mean-j-expression}, which is a function only of the potential
(through $\Omega_r$). Equation \eqref{eq:mean-j-expression} does not
exhaust the possible tests for spherical ergodicity, but it is
sufficient for our present exploratory purposes.

For smooth potentials, $\Omega_r(E,j)$ varies very little between
$j=0$ and $j=j_c$, and one can approximate it very well as a function
of $E$ alone (Appendix \ref{sec:Jr-j-halos}). In this case, the integrals follow analytically and one has the result
\begin{equation}
\langle j \rangle(E) \simeq \frac{2}{3} j_c(E)\textrm{.}\label{eq:mean-j-approx}
\end{equation}
This is a helpful simplification to set expectations, but throughout this paper when showing the spherically ergodic limit, we will use the exact expression given by equation \eqref{eq:mean-j-expression}.

\section{Example applications}\label{sec:ROI-connection}

So far we have motivated and derived a formal result that aspherical
systems are most stable when they appear ergodic in spherical
coordinates. We derived one practical consequence for the angular
momentum distribution, equation (\ref{eq:mean-j-expression}), which in an
approximate sense states that the velocity distribution will appear
isotropic. We are now in a position to test whether numerical
simulations actually tend towards this maximally-stable limit in a
variety of situations, beginning with cosmological collisionless dark
matter halos.

\subsection{Cosmological dark matter halos}\label{sec:cosmo-halos}

Let us re-examine the three high-resolution, dark-matter-only zoom
cosmological simulations used in the analysis of
\cite{2013MNRAS.430..121P}. The three each have several million
particles in their $z=0$ halos which correspond in turn to a dwarf
irregular, $L_{\star}$ galaxy and cluster. The force softening
$\epsilon$, virial radius $r_{200}$ (at which the mean density
enclosed is 200 times the critical density) and virial masses
$M_{200}$ are $65$, $170$, $690\,\pc$; $98$, $301$, $1430\,\kpc$ and
$2.8\times 10^{10}$, $8.0\times 10^{11}$, $8.7 \times 10^{13}\,\Msol$
respectively. For further details see \cite{2013MNRAS.430..121P}.

\begin{figure}
\includegraphics[width=0.49\textwidth]{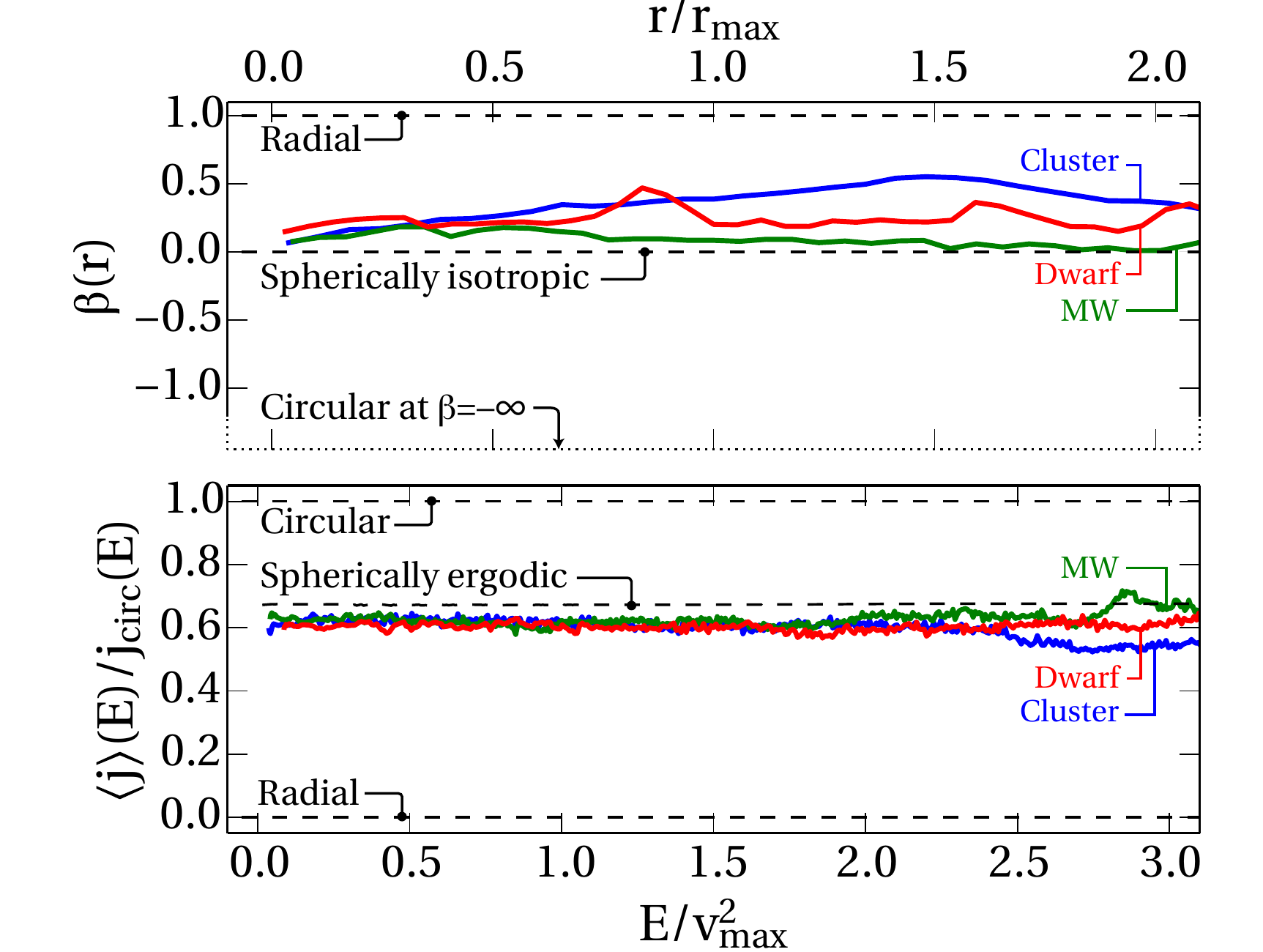}
\caption{The velocity anisotropy of the inner parts of three sample
  high-resolution cosmological dark matter halos (simulated without
  baryons), plotted as a function of radius (upper panel) and in
  energy shells (lower panel). The upper panel shows the classic
  velocity anisotropy $\beta(r)$ defined in the text, for which a
  purely radial population has $\beta=1$ and a population on circular
  orbits $\beta=-\infty$. The lower panel shows our alternative
  in energy space which can be more precisely related to the
  theoretical arguments presented in Section \ref{sec:most-stable-system}; here $\langle
  j \rangle (E)/\jcirc=0$ for radial orbits and $1$ for circular
  orbits, and approximately $2/3$ for an isotropic population.  Both panels show that the halos have near-isotropic orbits with a slight radial
  bias. The range of the two plots is roughly comparable, but we
  caution that the mapping from $r$ to $E$ is not unique (see Figure
  \ref{fig:r-to-E}). }\label{fig:cosmo-halos}
\end{figure}

The upper panel of Figure \ref{fig:cosmo-halos} shows the anisotropy
for our cosmological halos. To compare the three directly,
we scale the radius by $r_{\mathrm{max}}$ (respectively $27$, $57$ and
$340\,\kpc$), the radius at which the circular velocity
$(GM(<r)/r)^{1/2}$ reaches its maximum, $v_{\mathrm{max}}$ ($=56$,
$150$ and $610\,\kms$).  We restrict attention to the region well
within the virial radius; here, the anisotropy $\beta(r)$
typically lies between the purely radial and isotropic cases
\citep[e.g.][]{2008ApJ...685..739B,2010MNRAS.402...21N}.

We now want to link this relatively familiar velocity anisotropy to
the alternative $\langle j \rangle (E)$ statistic that was directly
predicted by the spherically ergodic property in Section
\ref{sec:most-stable-system}.  For each particle we calculate the
specific energy $E = \dot{\vec{x}}^2/2 + \Phi\left(|\vec{x}|\right)$,
where $\vec{x}$ is the vector displacement from the halo centre.  To
make this quantity agree exactly with $H_0$ in the terminology of
Section \ref{sec:most-stable-system}, we ignore asphericity when
calculating the potential energy, defining it as
\begin{equation}
\Phi(r) \equiv \int_0^r \frac{G M(<r')}{r'^2} \dd r'
\end{equation}
where $M(<r')$ is the mass enclosed inside a sphere of radius
$r'$. (The numerical integration is performed by binning particles in
shells of fixed width $\epsilon$, chosen to coincide with the force
softening in the simulation; within these bins the density is taken to
be constant.)  The physics is invariant if a constant is added to the
potential; we have chosen to fix its scale by setting $\Phi(0)=0$.

\begin{figure}
\includegraphics[width=0.49\textwidth]{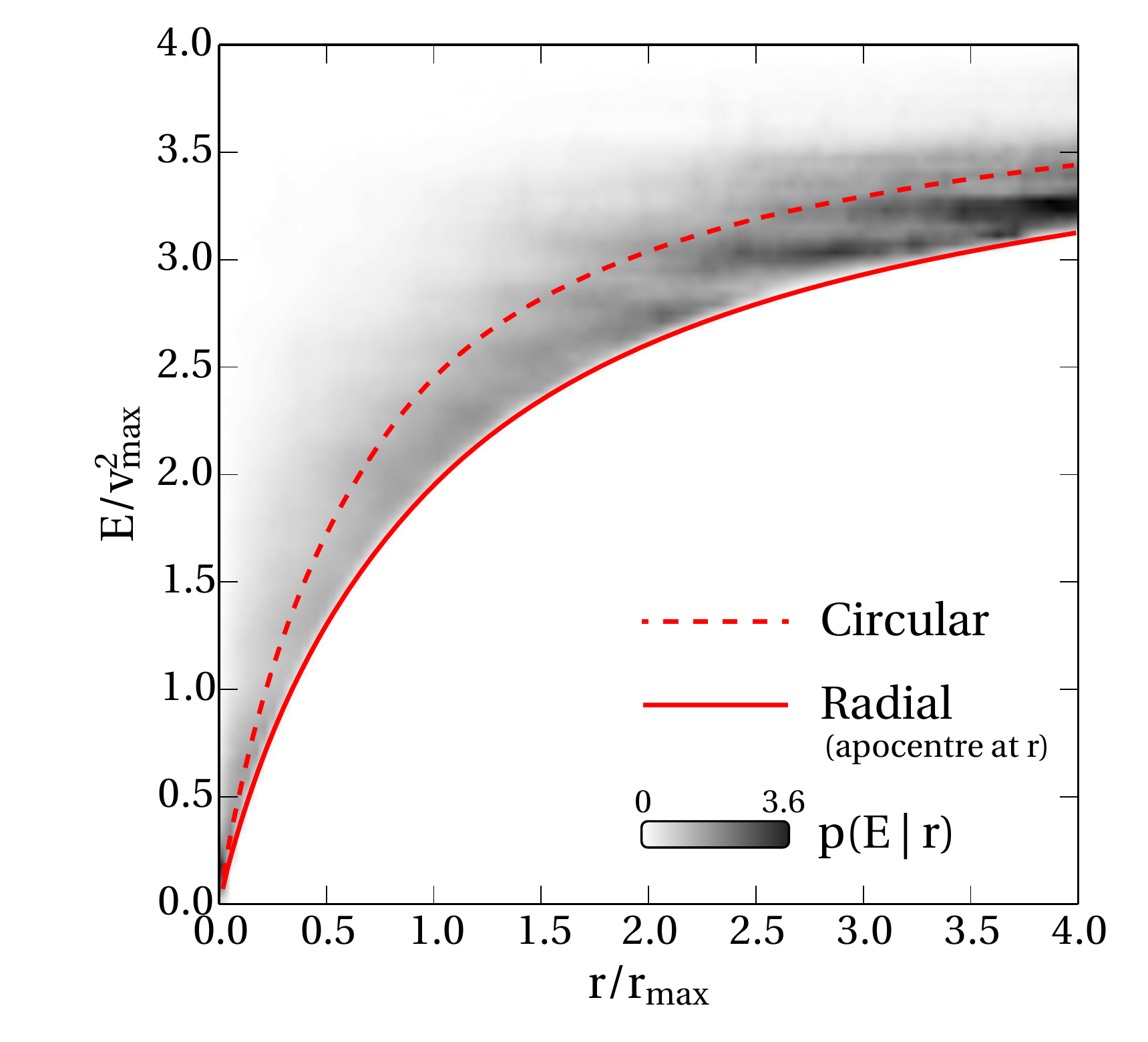}
\caption{The relationship between $r/\rmax$ and $E/\vmax^2$ for
  circular orbits (dashed line), radial orbits at apocentre (solid
  line) and particles in the `MW' run (density shows the number of
  particles with energy $E$ at each radius $r$). The
  mapping between energy and radius is fuzzy, so that anisotropy
  at high $E$ can easily contaminate $\beta(r)$ at small
  $r$.}\label{fig:r-to-E}
\end{figure}

For each particle we also calculate the specific angular momentum $j=\left| \vec{x} \times \vec{\dot{x}}
\right|$, and  the specific angular momentum of a circular
orbit at the same energy, $j_{\mathrm{circ}}(E)$ which is given by
simultaneously solving
\begin{align}
E &= \Phi(r)+\frac{\jcirc^2}{2 r^2} \hspace{0.5cm}\textrm{  and  }\hspace{0.5cm} \Phi'(r) = \frac{\jcirc^2}{r^3}\textrm{,}
\end{align}
to eliminate $r$ in favour of $\jcirc$.

We plot $\langle j \rangle(E)/\jcirc(E)$ in bins containing 1\,000 particles each in the lower panel of Figure \ref{fig:cosmo-halos}. To facilitate comparison with the top panel, a population on purely radial orbits would have $\beta=1$ and $\langle j \rangle/\jcirc=0$, whereas a purely circular distribution function corresponds to $\beta=-\infty$ or $\langle j \rangle/\jcirc=1$. Isotropic, purely spherical populations have $\beta(r)=0$ and $\langle j \rangle/\jcirc \simeq 2/3$ as discussed at the end of Section \ref{sec:sphergodic-predictions}. When compared against each other in this way, the two panels agree well: the populations are on near-isotropic orbits with a slight radial bias.

Quantitatively, how well do these results agree?
As a guideline, we can compare results for models with constant anisotropy. From \cite{BinneyTremaine2008}, a distribution function $f(j,E)=j^{-2\beta} f_1(E)$ generates a constant anisotropy $\beta(r)=\beta$. We can calculate the connection to the new statistic by generalising the reasoning of Section~\ref{sec:sphergodic-predictions}, with the result that
\begin{equation}
\langle j \rangle(E) = \frac{\int_0^{\jcirc} \dd j \, \Omega_r^{-1} \, j^{2-2\beta}}{\int_0^{\jcirc} \dd j \, \Omega_r^{-1} j^{1-2\beta}} \simeq \frac{2-2\beta}{3-2\beta} \jcirc\textrm{,}\label{eq:beta-to-j}
\end{equation}
where the first result is exact and the second follows from assuming $\Omega_r$ is independent of $j$ (which is an excellent approximation; see Appendix \ref{sec:Jr-j-halos}). Consistent with equation \eqref{eq:mean-j-approx}, $\beta=0$ gives $\langle j \rangle(E)/\jcirc \simeq 2/3$. But as the system becomes more radially biased we can now calculate that, for example, $\beta=0.2$ corresponds to $\langle j \rangle(E)/\jcirc \simeq 0.62$. Despite the various approximations involved, these values therefore correctly relate the values of $\beta$ in the top panel of Figure~\ref{fig:cosmo-halos} with the $\langle j \rangle$ results in the lower panel.

That said, a detailed comparison as a function of radius is hard
because particles at a given radius $r$ have a wide spread of energies
$E$. Figure \ref{fig:r-to-E} illustrates this relationship for the
`MW' halo. The density shows the probability of a particle at radius
$r$ also having specific energy $E$, $p(E|r)$. The minimum $E$ at each $r$ is set by $\Phi(r)$, which gives the energy of a particle at apocentre (solid line). A more typical $E$ is given by the energy of a circular orbit at $r$ (dashed line), and this gives some intuition for mapping results from the top panel of Figure~\ref{fig:cosmo-halos} onto the bottom panel. However, any $E$ exceeding $\Phi(r)$ is theoretically possible. So $\beta$ at any given radius actually represents an average over particles of many different energies.

\subsection{The classical radial orbit instability}\label{sec:classic-roi}

\begin{figure}
\includegraphics[width=0.49\textwidth]{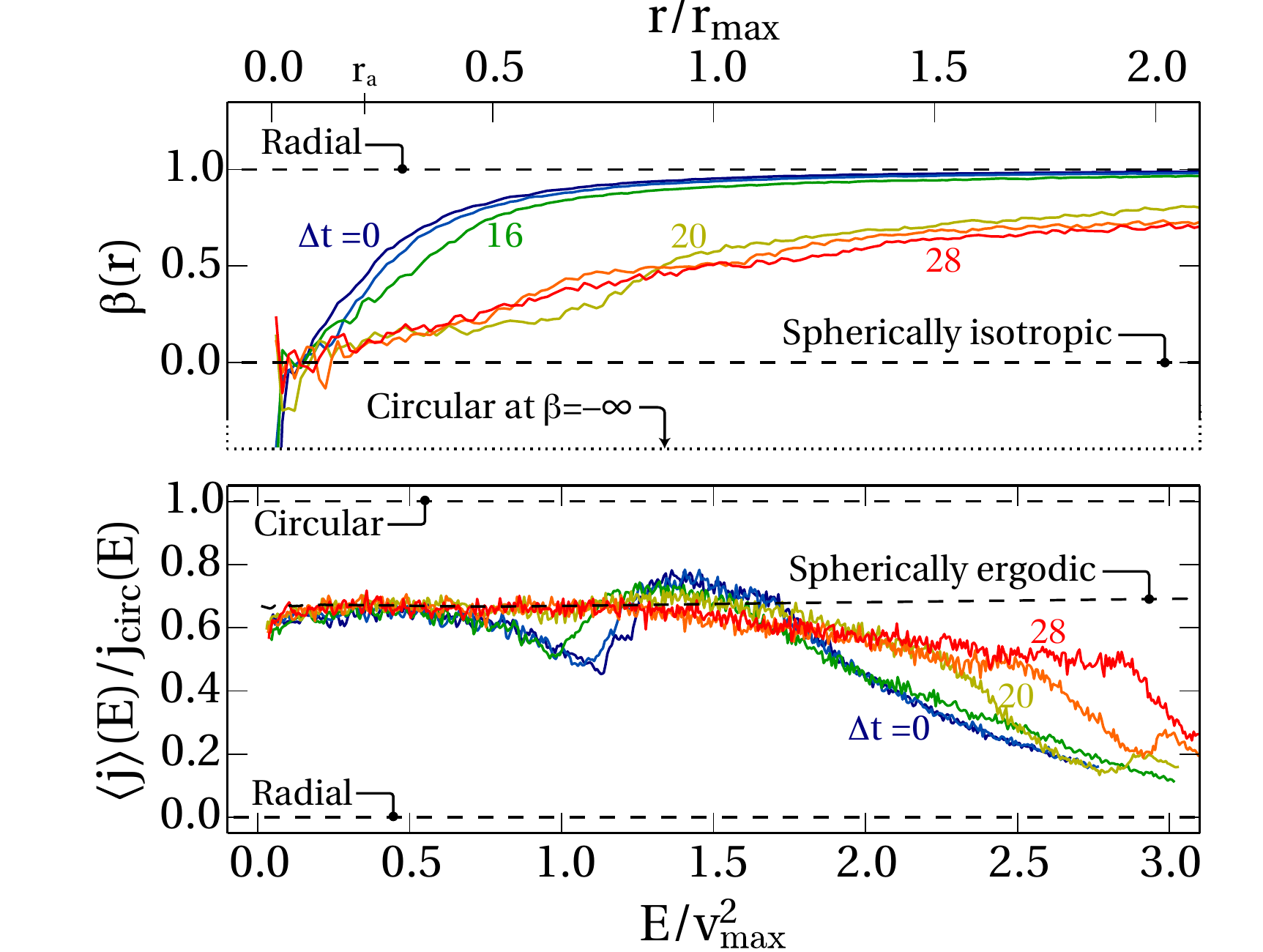}
\caption{The $\beta(r)$ velocity anisotropy (upper panel) and its
  equivalent in energy space $\langle j \rangle(E)$
  (lower panel, as Figure \ref{fig:cosmo-halos}) for the
  evolution over time of a halo that is initially in equilibrium, but
  unstable against the radial orbit instability. The initial
  conditions are represented by the dark blue line, $t=0$; there is a
  delay of $12 \tdyn$ before any significant evolution can be seen
  (light blue line). Subsequent lines are shown every $4 \tdyn$. Once
  the instability kicks in and generates aspherical perturbations,
  there is a rapid evolution towards the predicted spherically
  ergodic limit (lower panel).}\label{fig:j_E_evolution}
\end{figure}

Having established that, loosely speaking, $\langle j \rangle(E)$
represents the anisotropy in energy shells in the same way that
$\beta(r)$ does in radial shells, we can return to our prediction
(Section~\ref{sec:sphergodic-predictions}) for the former quantity,
which is shown by the dashed line in the lower panel of Figure
\ref{fig:cosmo-halos}. The prediction is almost, but not quite,
satisfied in cosmological dark matter halos. Since the condition is
only reached in a maximally-stable object, approximate agreement
is an acceptable situation.

Because cosmological halos initially form from near-cold collapse, the
radial orbit instability
\citep{vanAlbada82ROI,Barnes86ROI,1991MNRAS.248..494S,Weinberg91} is
invoked to explain how the radially infalling material gets scattered
onto a wider variety of orbits
\cite{2006ApJ...653...43M,2008ApJ...685..739B,2009ApJ...704..372B},
isotropising the velocity dispersion. Most tellingly, numerical
experiments by \cite{1999ApJ...517...64H,2006ApJ...653...43M} show
that the suppressing the instability (by switching off non-radial
forces) results in a qualitatively different density profile as the
end-point of collapse. We will now show that the isotropisation of
velocity dispersion during the radial orbit instability can be
interpreted in terms of an evolution towards stability in the terms of
Section \ref{sec:most-stable-system}.

Consider what happens to a halo that is intentionally designed
to be unstable. First, we will show the classic radial orbit
instability (ROI) at work by constructing a spherical halo with
particles that are on radially-biased orbits. We initialise our
particles such that they solve the Boltzmann equation and so are
stable in exact spherical symmetry. In practice, however, the strong
radial bias means that any slight numerical noise will trigger the
ROI.  By initialising an unstable equilibrium in this way, we avoid
confusion from violent relaxation processes associated with
out-of-equilibrium collapse \citep{1967MNRAS.136..101L}.

Specifically, the initial conditions are set up in a similar fashion to
\citet{2006MNRAS.367..387R}, with particle positions drawn from a
generalised Hernquist density profile
\citep{Hernquist1990,1993MNRAS.265..250D}:
\begin{equation}
\rho(r) = \frac{M(3-\gamma)}{4\pi a^3}
\left(\frac{r}{a}\right)^{-\gamma}\left(1+\frac{r}{a}\right)^{\gamma-4}\textrm{,}
\label{eqn:hern}
\end{equation}
which has a circular velocity reaching a peak at $\rmax=(2-\gamma)a$ and implies the gravitational potential
\begin{equation}
\Phi(r) = \frac{GM}{2-\gamma}\left[\left(1+a/r\right)^{\gamma-2}-1\right]\textrm{.}
\end{equation}
We choose $\gamma=1$ to roughly mimic an NFW halo \citep{1997ApJ...490..493N} in the innermost parts of interest. The velocities are sampled (using an
accept-reject algorithm) from a numerically-calculated Osipkov-Merritt
distribution function \citep{BinneyTremaine2008}:
\begin{equation}
f(Q) = \frac{\sqrt{2}}{4 \pi^2} \frac{\dd}{\dd Q} \int_Q^0 \frac{\dd
  \Phi}{\sqrt{\Phi - Q}} \frac{\dd}{\dd \Phi} \left[\left(
  1+r(\Phi)^2/r_a^2\right) \rho(r(\Phi) \right]
\label{eqn:OM}
\end{equation}
where the parameter $Q$ is defined by
\begin{equation}
Q \equiv E + \frac{j^2}{2r_a^2}\textrm{.}
\end{equation}
The value of $r_a$ is known as the anisotropy radius because the
velocity anisotropy is given by
\begin{equation}
\beta(r) = \frac{1}{1 + r_a^2/r^2}\textrm{,}
\end{equation}
showing that $\beta(r) \simeq 0$ (isotropic) for $r \ll r_a$ and
$\beta(r) \simeq 1$ (radial) for $r \gg r_a$. We used the minimum
value of $r_a$ for
which the distribution function is everywhere positive, making the orbits as radially biased as possible; for $\gamma =
1$, this is $r_a \simeq 0.21$ \citep{1997ApJ...490..136M}.
We draw $10^6$ particles and evolve the system using RAMSES \citep{Teyssier02ramses}, with mesh refinement based on the number of particles per cell, resulting in a naturally adaptive force softening reaching a minimum of $\epsilon = 90\,\pc$.

Our expectation that numerical noise triggers the ROI is borne out by
the numerical experiment. The upper panel of Figure
\ref{fig:j_E_evolution} shows  the radial anisotropy $\beta(r)$ over
time. We have defined a single dynamical time, $\tdyn$, at the peak of
the velocity curve so that $\tdyn \equiv \rmax/\vmax$.
The six solid lines show the population at $t=0, 12, 16, 20, 24$ and $28$ dynamical times. At first, $\beta(r)$ appears stable, but suddenly after $16\,\tdyn$ it becomes considerably more isotropic. Over the same timescale, asphericity in the potential develops. To demonstrate this, we determine the inertia tensor of the entire density distribution and calculate the ratio of the principal axes; at $t=0$, the ratio is $1.0$ by construction. By $16\,\tdyn$ the ratio is $1.8$. It stabilises at around $26\,\tdyn$ with a ratio of $4.3$.

This is symptomatic of the classic radial orbit instability in action. We can follow the same process from our energy/angular-momentum standpoint in the bottom panel of Figure \ref{fig:j_E_evolution}. The initial conditions ($\Delta t = 0$) have a kink in them, with most regions of energy space appearing radially biased but some showing a slight circular preference (around $E \simeq 1.5 \vmax^2$). We verified that this is an artefact of the Opsikov-Merritt construction and is consistent with the radial $\beta(r)$ being radially biased everywhere; recall Figure \ref{fig:r-to-E} shows that the relationship between $r$ and $E$ is non-trivial.

As time progresses, the $\langle j \rangle(E)$ distribution correctly
tends towards the spherically ergodic (SE) limit, as predicted. The SE
limit is attained to good accuracy for energies $E<\vmax^2$ by $\Delta
t = 20 \,\tdyn$. At larger energies, it is likely difficult to achieve
SE because of the long time-scales and weak gravitational fields
involved.  Like $\langle j \rangle$, $\beta(r)$ at $28\,\tdyn$ shows
an isotropic distribution at the centre ($r \to 0$); but $\beta(r)>0$
everywhere for $r>0.1\,\rmax$. We verified that this continuing radial
bias throughout the halo is produced by a number of high-energy,
loosely-bound particles plunging through, i.e. the high-$E$ particles
from the lower panel in Figure \ref{fig:j_E_evolution} pervade all
radii in the upper panel.

Let us briefly recap: what we have so far is a new view of an existing
phenomenon. We have recast the ROI and its impact on
spherically-averaged quantities as an evolution towards an
analytically-derived maximally stable class of distribution
functions. Viewed in energy shells instead of radial shells, the
distinction between the regions that reach the SE limit and the
slowly-evolving, loosely-bound regions that remain radially biased is
considerably clearer.  Now, because our underlying analytic result is
derived without requiring the population to be self-gravitating, we
can now go further and consider a different version of the radial
orbit instability -- and show that it continues to operate even when a
system is in a completely stable equilibrium.

\subsection{The continuous radial orbit instability}\label{sec:cont-roi}

\begin{figure}
\includegraphics[width=0.49\textwidth]{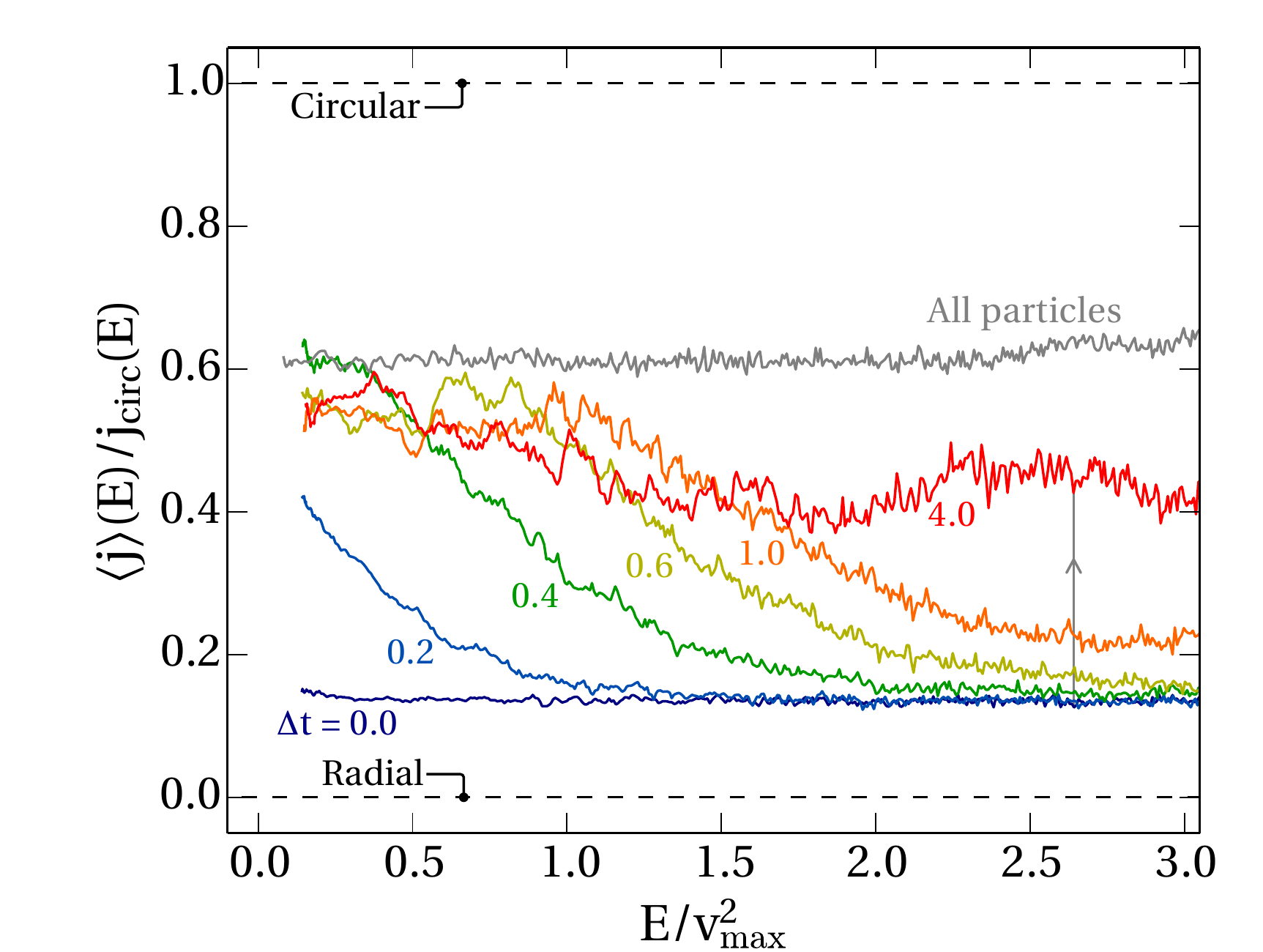}
\caption{The mean angular momentum of particles as a function of energy, like the lower panel of Figure \ref{fig:j_E_evolution} -- but now for a subpopulation in a dark matter halo that is globally in equilibrium. At $\Delta t=0$, we select particles on
  predominantly radial ($j/\jcirc<0.2$) orbits; at later times, the subpopulation mean evolves
  back towards the population mean. The lines from bottom to top show
  the state at selection and after $0.2$, $0.4$, $0.6$, $1.0$ and
  $4.0\, \tdyn$ respectively.
}\label{fig:subpopulation}
\end{figure}

Our next target for investigation is to broaden the conditions:
according to the results of Section \ref{sec:most-stable-system}, a
subpopulation of particles should undergo something much like the
radial orbit instability even when the {\it global} potential is
completely stable. A suitable name for this phenomenon would seem to
be the ``continuous radial orbit instability", since it continues
indefinitely after the global potential has stabilised.

We perform another numerical experiment to demonstrate the
effect. First, to avoid confusion from cosmological infall and tidal
fields, we create a stable, isolated, triaxial cosmological halo by
extracting from our cosmological run a region of $3 r_{200}$ around
our `Dwarf'. We then evolve this isolated region for $2\,\Gyr$ to
allow any edge effects to die away, and verify that the density
profile out to $r_{200}$ is completely stable. As before we define a dynamical
time for the system of $t_{\mathrm{dyn}} \equiv
r_{\mathrm{max}}/v_{\mathrm{max}} = 470\,\Myr$.
%
%

After the $2\,\Gyr \simeq 4 \tdyn$ has elapsed, we select all
particles with $j<0.2\,j_{\mathrm{circ}}$. These particles are, at the
moment of selection, on preferentially radial orbits.  We  trace
our  particles forward through time, measuring $\langle j \rangle(E)$ in each snapshot. The results are shown in Figure
\ref{fig:subpopulation} for various times between $\Delta t = 0.0$ and
$4.0\,t_{\mathrm{dyn}}$ after selection. Over this period, the mean
angular momentum significantly increases towards the spherically ergodic limit at every energy. The changes are
much faster at low energies where the particles are tightly bound and
the local dynamical time is short compared to the globally-defined $\tdyn$.

The evolution is rapid until $4\,\tdyn$ after which the $\langle j
\rangle(E)$ remains near-static (except at $E>2.5 \vmax^2$ where it
continues to slowly rise). Eventually the subpopulation has $\langle j
\rangle(E) \simeq 0.5 \jcirc(E)$ independent of $E$. This establishes that particles at low angular momentum within a completely
stable, unevolving halo, automatically evolve towards higher angular
momentum. After a few dynamical times, their angular momentum becomes
comparable to that of a randomly-selected particle from the full
population, although with a continuing slight radial bias.

One can understand this incompleteness in a number of equivalent
ways. Within our formal picture, it arises from the fact that only
certain $\delta H_{\vec{n}}$ are non-zero in equation
\eqref{eq:aspherical-evolution}. More intuitively, the continuing
process of subpopulation evolution is being driven by particles on box
orbits that change their angular momentum at near-fixed energy -- but
not all particles are on such orbits, and so some memory of the
initial selection persists.

\subsection{Observations of dwarf spheroidal galaxies}\label{sec:dsph}

\begin{figure}
\includegraphics[width=0.49\textwidth]{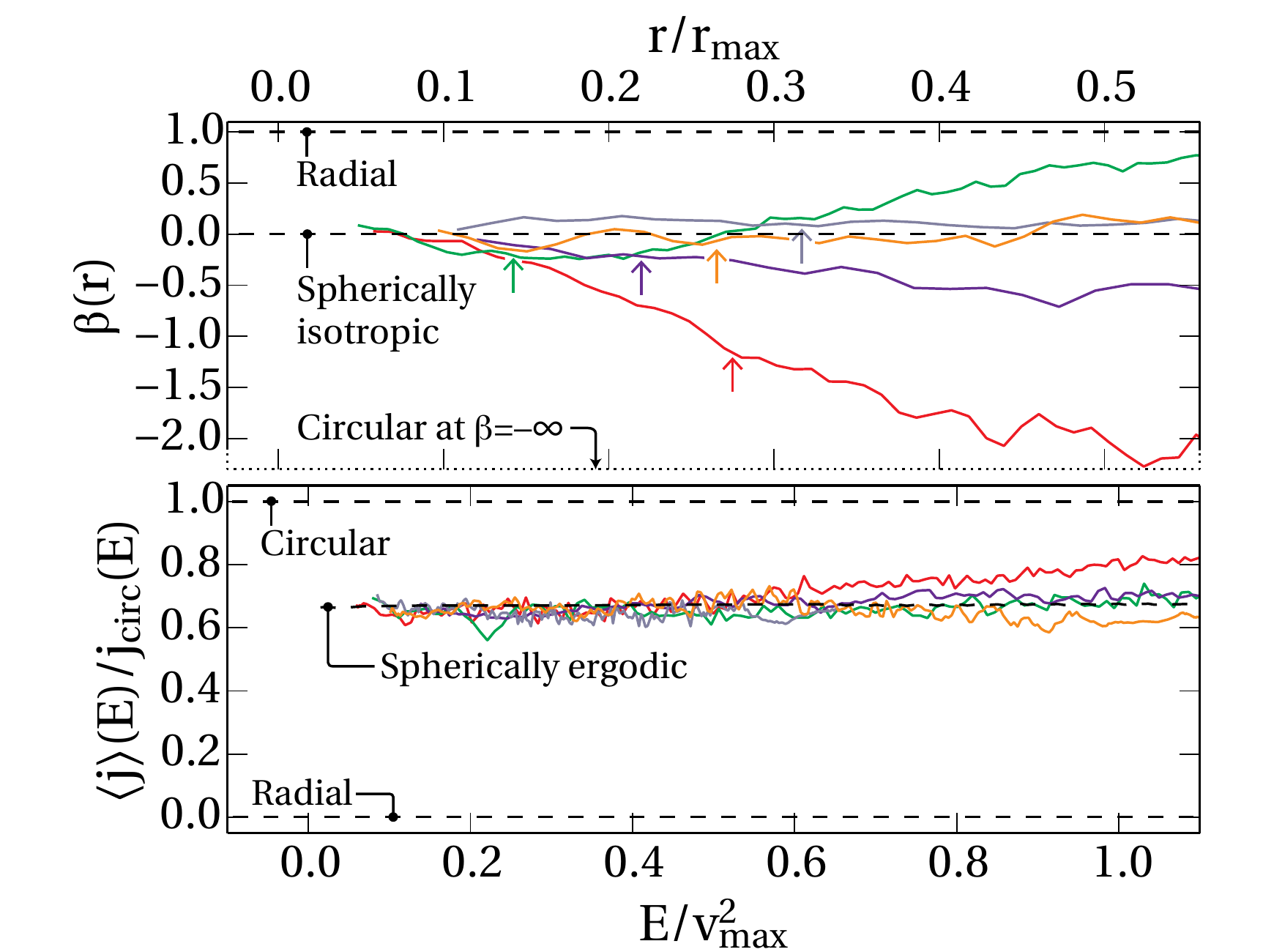}
\caption{The $\beta(r)$ (upper panel) and $\langle j \rangle(E)$
  (lower panel) relations for the stellar populations of dwarf
  satellites around a Milky-Way-like central. In the upper panel
  arrows indicate the location of $r_{1/2}$, the radius enclosing half
  the stellar mass. The profile is plotted exterior to $3\epsilon$
  where $\epsilon$ is the softening length. Because of the tidal
  interaction with the parent galaxy, the outer parts of each object
  are out of equilibrium and display biases ranging from strongly
  radial to circular. The lower panel shows the same populations in
  energy space. At low energies, tightly bound particles are now seen
  to be close to the spherically ergodic limit. At higher energies, a
  spread is seen but not as large as would naively be expected from
  the $\beta(r)$ relation. }\label{fig:stellar}
\end{figure}

In the previous section, we established that subpopulations on
initially radially-biased orbits evolve towards velocity isotropy even
if the global potential is stable. We explained this in terms of our
earlier calculations, and now turn to how the results might impact on
observations.

An important current astrophysical question is how dark matter is distributed within low-mass galaxies: this can discriminate between different particle physics scenarios \citep{pontzen2014nature}. The smallest known galaxies, dwarf spheroidals surrounding our Milky Way, in principle provide a unique laboratory from this perspective. But determining the distribution of dark matter is a degenerate problem because of the unknown transverse velocities of the stellar component \cite[e.g.][]{Charbonnier11}. Furthermore the use of spherical analyses seems inappropriate since the underlying systems are known to be triaxial \cite[e.g.][]{Bonnivard15}.

This is exactly the kind of situation we set up in our initial
calculations (Section \ref{sec:most-stable-system}): an aspherical
system being analysed in spherical symmetry. We have shown that the
results apply both to tracer and self-gravitating populations. So, we
can go ahead and apply it to the stars in observed systems: the
stellar population of a dwarf spheroidal system will be maximally
stable if it appears ergodic in a spherical analysis. This implies a
strong prior on what spherical distribution functions are actually
acceptable and therefore, in principle, lessens the anisotropy
degeneracy.

This idea warrants exploration in a separate paper; here we will briefly test whether the idea is feasible by inspecting some simulated dwarf spheroidal satellite systems. In particular we use a gas-dynamical simulation of ``MW" region (see above) using exactly the same resolution and physics as \cite{2012arXiv1207.0007Z}; see that work for technical details (although note the actual simulation box is a different realisation).

We analyse all satellites with more than $10^4$ star particles, which gives us objects lying in the ranges $5.4 \times 10^7<M_{\star}<4.9 \times 10^8\,\Msol$, $28<\vmax<60\,\kms$ and $4.6<\rmax<9.6\,\kpc$. We first verified that these do not host rotationally supported disks; we then calculated $\beta(r)$ profiles between $3 \epsilon \simeq 0.5\,\kpc$ and $0.6\,\rmax$. The stellar half-light radius is $0.15 < r_{1/2}/\rmax < 0.3$, so our calculations extend into the outer edges of each visible system.

The results are shown in the upper panel of Figure \ref{fig:stellar}, with arrows indicating $r_{1/2}$. The outer regions display a range of different behaviours from strongly radial to circular. However, in each case the $\beta(r)$ is nearly isotropic as $r\to 0$. This is consistent with a view in which the centres have achieved stability while the outer parts are being harassed by the parent tides and stripping.

The picture is reinforced by considering the $\langle j \rangle(E)$
statistic (lower panel of Figure \ref{fig:stellar}). Here, stars on
tightly bound orbits (small $E$) are very close to the spherically
ergodic limit. In the less-bound regions, the agreement worsens. However a quantitative analysis using the approximate relation \eqref{eq:beta-to-j} shows that the $\langle j \rangle(E)$ relation stays much closer to the spherically ergodic limit than the $\beta(r)$ relation naively implies. This is probably because the shape of $\beta(r)$ is in part determined by out-of-equilibrium processes, coupled to the multivalued relationship between $r$ and $E$ (Figure \ref{fig:r-to-E}).

In any case these results suggest that we should adopt priors on
spherical Jeans or Schwarzschild analyses that strongly favour near-isotropy in the centre of spheroidal systems -- or, more accurately, strongly favour near-ergodicity for tightly bound stars.  In future work we will expand on these ideas and apply them to observational data, since they could lead to a substantial weakening of the problematic density-anisotropy degeneracies.

\subsection{Dark matter cusp -- core transitions}\label{sec:cusp-core}

\begin{figure}
\includegraphics[width=0.49\textwidth]{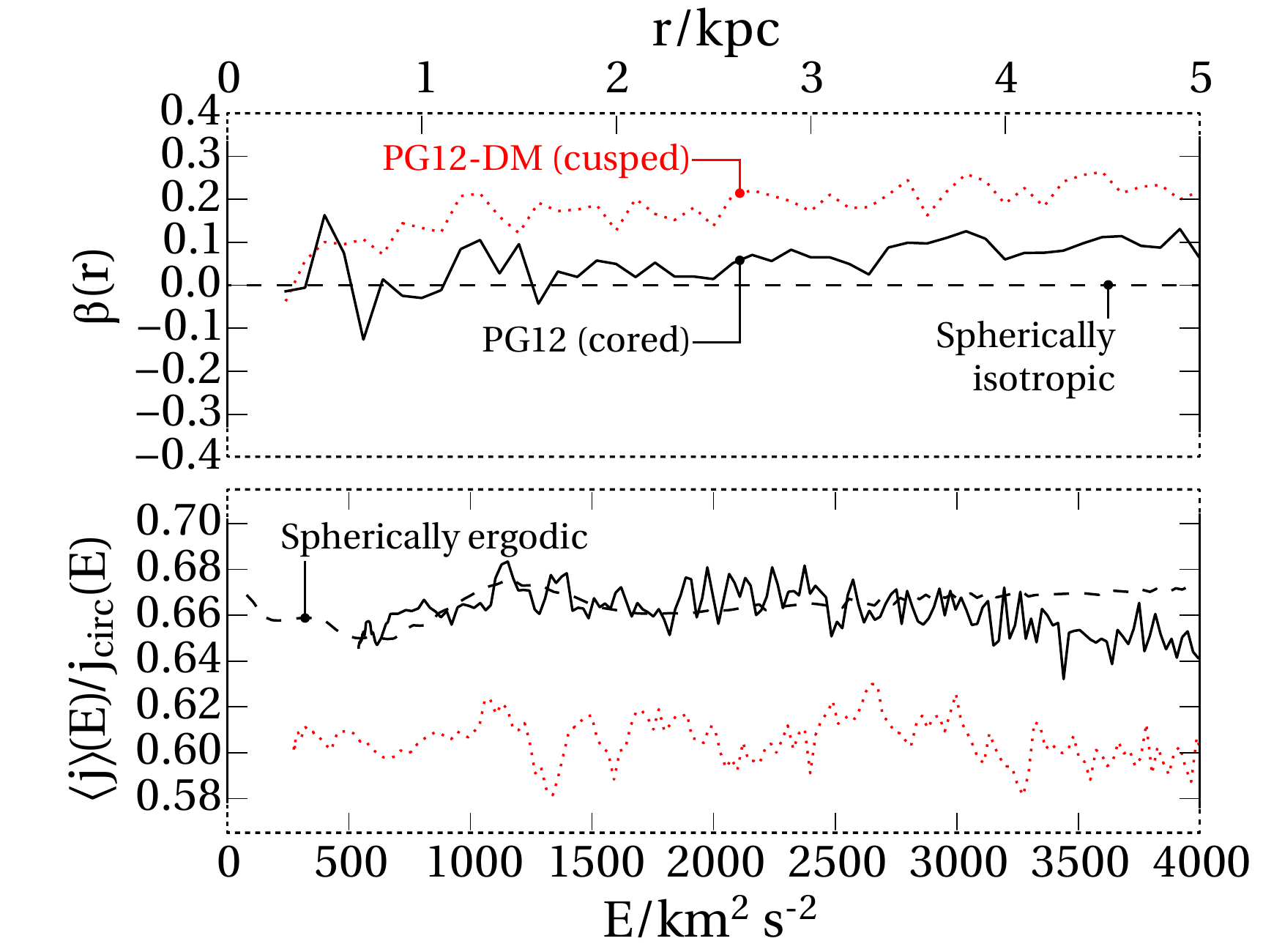}
\caption{The velocity anisotropy ($\beta(r)$, upper panel) and angular
  momentum ($\langle j \rangle(E)$, lower panel) for cusped and cored
  halos from PG12; note the much-expanded $y$-axis scales relative to
  previous figures, which are required to highlight the
  differences. The cored cases (solid lines) are almost perfectly
  spherically ergodic (lower panel) and hence isotropic (upper
  panel). This contrasts with the cusped case (dotted lines) which has
  a slight but significant radial bias (seen as high $\beta$ in the
  upper panel and low $j$ in the lower
  panel). }\label{fig:jE_cusp_v_core}
\end{figure}

In \citet[][henceforth PG12]{2012MNRAS.421.3464P} we established that
dark matter can be redistributed when intense, short, repeated bursts
of star formation repeatedly clear the central regions of a forming
dwarf galaxy of dense gas \citep[see
also][]{2005MNRAS.356..107R,2006Natur.442..539M}. The resulting
time-changing potential imparts net energy to the dark matter in
accordance with an impulsive analytic approximation. This type of
activity has now been seen or mimicked in a large number of simulations, allowing
glimpses of the dependency of the process on galactic mass, feedback
type and efficiency
\citep{2012MNRAS.422.1231G,2012ApJ...759L..42P,2011arXiv1112.2752M,teyssier13,2012arXiv1207.0007Z,GK13,2013arXiv1306.0898D,2015arXiv150202036O}.

We based our PG12 analysis on 3D zoom cosmological simulations, but our
analytic model assumed exact spherical symmetry. By definition, in
the analytic model all particles conserve their angular momentum at
all times. Taken at face value, energy gains coupled to constant
angular momentum would leave a radially biased population in the
centre of the halo.

The top panel of Figure
\ref{fig:jE_cusp_v_core} shows the measured velocity anisotropy in the inner
$5\,\kpc$ for the zoom simulations in PG12 at $z=0$; the solid line
shows the feedback simulation which has developed a core, whereas the
dotted line shows the dark-matter-only simulation which maintains its
cusp. The difference between the two cases is the opposite
to that naively expected: the centre of the cored halo has a more
isotropic velocity dispersion than the centre of the cusped halo.
The lower panel shows the equivalent picture in energy space (noting
that
$\Phi(0)=0$ and $\Phi(5\,\kpc)=3\,790\,\mathrm{km^2\,s^{-2}}$). For
clarity only a small fraction of the $0<j/\jcirc<1$ interval is
now plotted; the differences between the cusped (dotted) and cored (solid)
lines are relatively small, but significant. At all energies the cored
profile has more angular momentum than the cusped profile; in fact, it
lies right on the spherically-ergodic limit (dashed line) whereas
the cusped profile is biased by around $0.06$ to radial
orbits\footnote{The dashed line for the isotropic $j/\jcirc$ in the
  lower panel is calculated using the cored potential. However the
  dependence on potential is very weak, as discussed in Section
  \ref{sec:ROI-connection}; calculating it instead with the cusped
  potential leads to differences of less than one percent at every
  energy.}.

While the PG12 model correctly describes the energy shifts, it misses
the angular momentum aspect of core creation, which has been
emphasised elsewhere \citep[e.g.][]{2006ApJ...649..591T}. We will now show that
a complete description of cusp--core transitions involves two
components: energy gains consistent with PG12, and re-stabilisation of
the distribution function consistent with the description of Sections
\ref{sec:population} and \ref{sec:cont-roi}. We again use the RAMSES
code to follow isolated halos with particle mass $2\times
10^4\,\Msol$; however we adapted the code to add an external,
time-varying potential to the self-gravity. 

\begin{figure}
\includegraphics[width=0.49\textwidth]{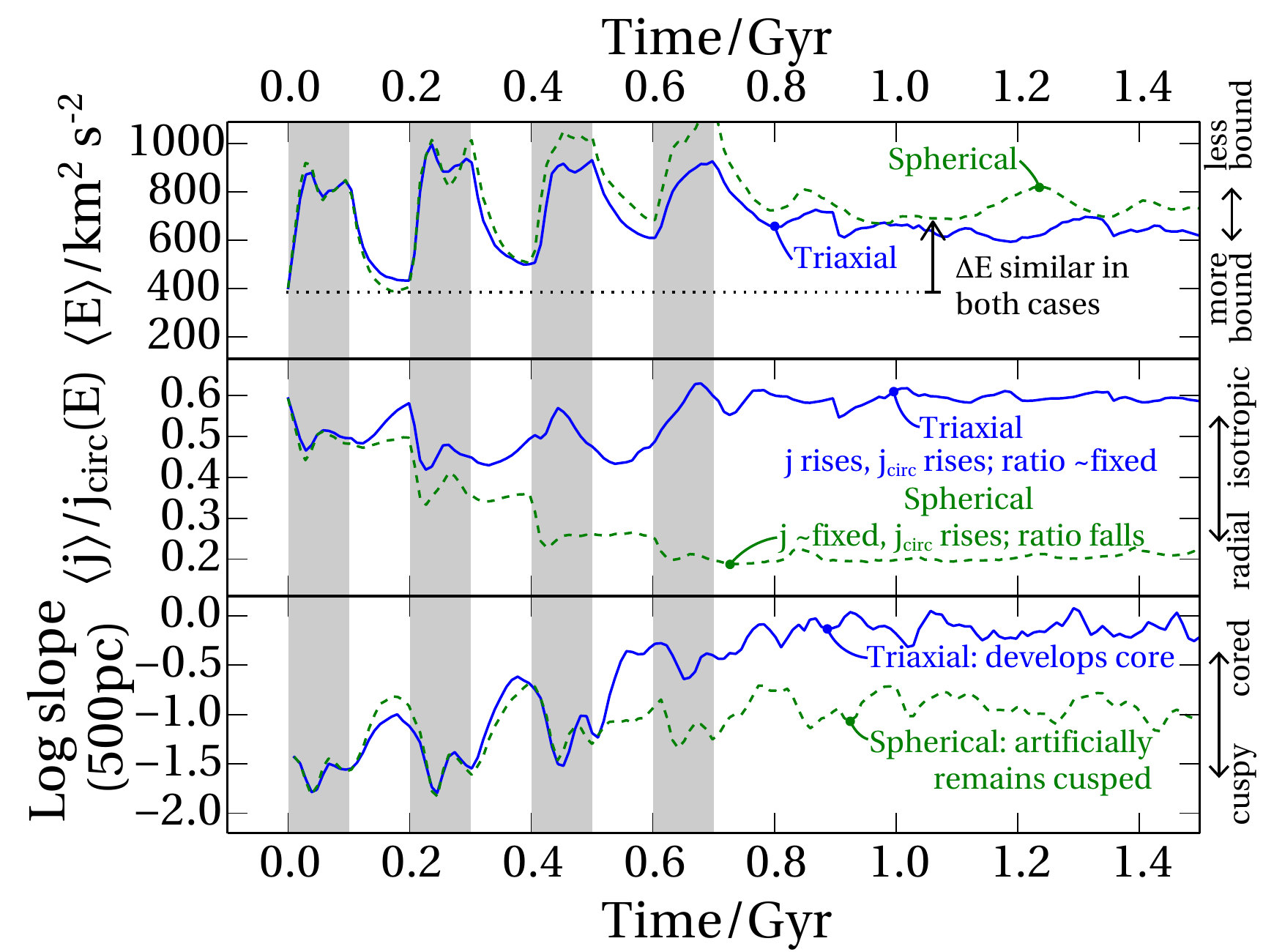}
\caption{The PG12 mechanism is reproduced by adding an external
  potential (to mimic `gas') to a DM-only simulation for the time
  intervals indicated by the grey bands. We measure the response of a
  completely spherical halo (dotted lines) and a realistically
  triaxial halo (solid lines) as described in the text. In both cases
  we track the mean energy (top panel) and angular momentum (middle
  panel) of the $0.1\%$ of particles that start out most bound. We
  also measure the log slope of the density profile at each time
  output (bottom panel). The energy shift (top) predicted by PG12 is
  found to hold in both cases. In the spherical case, the angular
  momentum remains constant up to discreteness effects; it therefore
  drops relative to the circular angular momentum (middle
  panel). Conversely, in the triaxial case, the continuous radial
  orbit instability causes the angular momentum to rise in proportion
  to $j_{circ}$. Only when the angular momentum is allowed to rise
  does a significant core form, meaning that the
  spherical case unphysically suppresses core formation (lower
  panel). }\label{fig:cuspflattening}
\end{figure}

We took our stabilised, isolated, triaxial dark matter halo extracted
as described in Section \ref{sec:cont-roi} and created a sphericalised
version of it as follows. For each particle we generate a random
rotation matrix following an algorithm given by \cite{graphicsgems},
then multiply the velocity and position vector by this
matrix. Finally, we verified that the final particles are distributed
evenly in solid angle, and that the spherically-averaged density
profile and velocity anisotropy is unchanged. The
triaxial and spherical halos are both NFW-like and stable over more than
$2\,\Gyr$ when no external potential is applied.

For our science runs, we impose an external potential corresponding to
$10^8\,\Msol$ gas in a spherical ball $1\,\kpc$ in radius, distributed
following $\rho\propto r^{-2}$; this implies a potential perturbation
of $\Delta \Phi = -700\,\km^2\,\s^{-2}$ at $500\,\pc$, for
instance. The potential instantaneously switches off at $100\,\Myr$,
back on at $200\,\Myr$, off at $300\,\Myr$ and so forth until it has
accomplished four ``bursts''. The period and the mass in gas is
motivated by Figures~1 and~2 of PG12 and Figure~7 of T+13. We also
tried imposing potentials with different regular periods and with
random fluctuations, none of which altered the behaviour described below.

The top two panels of Figure~\ref{fig:cuspflattening} show the
time-dependent behaviour of the central, most-bound $0.1\,\%$ of
particles. The triaxial and spherical halo results are illustrated by
solid and dashed lines respectively. The top panel shows how the
coupling of the external potential is similar; in particular, it
results in a mean increase in specific energy of $\Delta E \simeq
200\,\km^2\,\s^{-2}$ for particles in both cases. The final shift in
the spherical case is very slightly larger than that in the triaxial
case, a difference which is unimportant for what follows (it would
tend to create a larger core if anything).

The middle panel displays the mean specific
angular momentum $j$ of the same particles as a fraction of the
circular angular momentum $\jcirc(E)$. In the spherical case (dashed
line), this quantity drops significantly because $j$ is fixed but
$\jcirc$ is rising over time (since it increases with $E$). By
contrast in the triaxial case $j/\jcirc$ returns to its original
value, meaning that $j$ has risen by the same fraction as
$\jcirc(E)$. This is dynamic confirmation of the discussion earlier in
this Section: the spherically symmetric approach predicts an
increasing bias to radial orbits -- whereas in the realistic
aspherical case, the stability requirements quickly erase this bias.

The bottom panel of Figure \ref{fig:cuspflattening} shows the measured
density slope $\alpha = \dd \ln \rho / \dd \ln r$ at $500\,\pc$ for
the two experiments. Both start at $\alpha \simeq -1.5$; the triaxial
case correctly develops a core (with $\alpha\simeq -0.1$) whereas the
completely spherical case maintains a cusp (with $\alpha \simeq
-0.9$). This discrepancy is causally connected to the angular momentum
behaviour: the mean radius of a particle increases with increasing
angular momentum $j$, even at fixed energy $E$, so a typical particle
migrates further outwards in the triaxial case compared to the
spherical.

We can therefore conclude that asphericity is a pre-requisite for
efficient cusp--core transitions. However there is a subsidiary issue
worth mentioning. After about $3\,\Gyr$ we find that the spherical
halo autonomously does start increasing $\langle j \rangle / \jcirc$
and the dark matter density slope becomes shallower. This is because
the potential fluctuations have generated a radially biased population
which is unstable, and a global radial orbit instability is triggered
by numerical noise over a sufficiently long period (as in
Section \ref{sec:classic-roi}).

In fact, with `live' baryons rather than imposed fluctuations, there
are aspherical perturbations which accelerate the re-equilibration
process further and renew a global radial orbit instability (Section
\ref{sec:classic-roi}), encouraging the entire population towards
spherical ergodicity. As we saw in Figure \ref{fig:jE_cusp_v_core},
the cored halo from PG12 has an almost perfectly spherically-ergodic
population for $E<3\,000\,\km^2\,\s^{-2}$ (unlike the cusped case). We
verified that this is also true of T+13. Generating a core through
potential fluctuations seems to complete a relaxation process that
otherwise freezes out at an incomplete stage during collisionless
collapse.

\section{Conclusions}\label{sec:conclusions}

Let us return to the original question: how much do inaccuracies
inherent in the spherical approximation really matter in practical
situations? The answer is, unfortunately, that ``it depends''; but we
can now distinguish two cases as a rule of thumb:
\begin{enumerate}[1)]
\item For dynamical calculations or simulations, the inaccuracy
  matters a great deal. Neglecting the aspherical part of the
  potential unphysically freezes out the radial orbit instability and
  related effects, so can lead to qualitatively incorrect behaviour.
\item For the {\it analysis} of observations or simulations in
  equilibrium, the assumption is far more benign -- it is, in fact,
  extremely powerful when handled with care. The underlying aspherical
  system and the fictional spherical system both appear to be in
  equilibrium; the mapping between the two views yields striking insight
  into (for example) spheroidal stellar distribution functions and
  dark matter halo equilibria.
\end{enumerate}


These conclusions are based on the fact that, when aspherical systems
are analysed in spherical coordinates, there is an attractor solution
for the spherically-averaged distribution function $f_0$ -- namely, it
tends towards being ergodic (i.e. $f_0$ is well-approximated as a
function of energy alone). We demonstrated this using Hamiltonian
perturbation theory (see Section \ref{sec:most-stable-system} and
Appendix \ref{sec:f0-evolution}), and subsequently used the term
``spherically ergodic'' (SE) to describe a distribution function $f$
with the property that its spherical average $f_0$ is ergodic in this
way.

The result follows because the orbits do not respect spherical
integrals-of-motion such as angular momentum.  Note, however, that the
physical orbits may still possess invariants in a more appropriate set
of coordinates. The apparent chaotic behaviour and tendency of
particles to spread evenly at each energy is a helpful illusion caused
by adopting coordinates that are only partially appropriate.

In Section \ref{sec:ROI-connection} we showed that the idea has
significant explanatory power.  First, we inspected the selection of
equilibrium in triaxial dark matter halos (Section
\ref{sec:cosmo-halos}) which led us to consider the classical radial
orbit instability (Section \ref{sec:classic-roi}). We demonstrated
that the instability naturally terminates very near our SE limit
(lower panel, Figure \ref{fig:j_E_evolution}). Particles at high
energies have long dynamical times which causes them to freeze out:
they evolve towards, but do not reach, the SE limit. Because these
high-energy particles often stray into the innermost regions (Figure
\ref{fig:r-to-E}), the velocity anisotropy $\beta(r)$ continues to
display a significant radial bias at all $r$ after the instability has
frozen out.  Moreover in the case of self-consistently formed
cosmological halos (Figure \ref{fig:cosmo-halos}), even at low
energies there is a slight radial bias. The bias is only erased when a
suitable external potential is applied ({\it e.g.} during baryonic
cusp-core transformations), forcing the system to stabilise itself
against a wider class of perturbations than it can self-consistently
generate (Figure \ref{fig:jE_cusp_v_core}); we will return to this
issue momentarily.

One novel aspect of our analysis compared to previous treatments of
the radial orbit instability is that it applies as much to tracer
particles as to self-gravitating populations. As a first example, in
Section \ref{sec:cont-roi} we demonstrated how a subset of dark matter
particles chosen to be on radially-biased orbits mix back into the
population (Figure \ref{fig:subpopulation}). This is the case even for
halos that, as a whole, are in stable equilibrium; therefore we
referred to the phenomenon as a `continuous' radial orbit instability.

The same argument implies that stars undergo the radial orbit
instability as easily as dark matter particles. In particular, without
a stable disk to enforce extra invariants, dwarf spheroidal galaxies
likely have stellar populations with a near-SE distribution
(Section \ref{sec:dsph}; Figure \ref{fig:stellar}). This provides a
footing on which to base spherical Jeans or Schwarzschild analyses of
observed systems: it implies an extra prior which can be formulated
loosely as stating that $\beta(r) \to 0$ as $r \to 0$. Such a prior
could be powerful in breaking the degeneracy between density estimates
and anisotropy \citep[e.g.][]{Charbonnier11}, In turn tightening
limits on the particle physics of dark matter
\citep{pontzen2014nature}.

As a final application, we turned to the question of the baryonic
processes that convert a dark matter cusp into a core (Section
\ref{sec:cusp-core}).  Angular momentum is gained by individual
particles during the cusp-flattening process
\citep{2006ApJ...649..591T} but our earlier work (especially PG12) has
focussed primarily on the energy gains instead.  In Figure
\ref{fig:cuspflattening} we see that, in spherical or in triaxial dark
matter halos, time-dependent perturbations (corresponding to the
behaviour of gas in the presence of bursty star formation feedback)
always lead to a rise in energy. However, in the spherical case the
mean angular momentum as a fraction of $\jcirc$ drops
because the angular momentum of each particle is fixed, whereas
$\jcirc$ rises with $E$. Only in the triaxial case does $\langle
j\rangle / \jcirc$ stay constant, indicating a re-isotropisation
of the velocities. We tied the increase in $\langle j
\rangle$ to the continuous radial orbit instability which always
pushes a population with low $\langle j \rangle / \jcirc$ back towards
the SE limit (Section \ref{sec:cont-roi}). The
consequence is that -- in realistically triaxial halos -- the final
dark matter core size will be chiefly determined by the total energy
lost from baryons to dark matter, with little sensitivity to details
of the coupling mechanism.

Although the PG12 model can be completed neatly in this way, other
analytic calculations or simulations based on spherical symmetry will
need to be considered on a case-by-case basis. In our
exactly-spherical test cases (Figure \ref{fig:cuspflattening}), core
development is substantially suppressed. This certainly cautions
against taking the results of purely spherical analyses at face
value. On the other hand, any slight asphericity is normally
sufficient to prevent the unphysical angular-momentum lock-up -- in
particular, we verified that the simulations in \cite{teyssier13} were
unaffected by this issue because their baryons settle into a flattened
disk. The analytic result is generic, so the exact shape and
strength of the asphericity is a secondary effect in determining the
final spherically-averaged distribution function $f_0$.

That said, to understand the way in which $f_0$ actually evolves
towards the SE limit (and perhaps freezes out before it gets there)
requires going to second order in perturbation theory, as shown in
Appendix \ref{sec:f0-evolution}. At this point it may also become
important to incorporate self-gravity, i.e. the instantaneous
connection between $\delta f$ and $\delta H$. This approach has been
investigated more fully elsewhere, leading to a different set of
insights regarding the onset of the radial orbit instability
\cite[e.g.][]{1991MNRAS.248..494S} as opposed to its end state. The
present work actually suggests that the radial orbit instability can
be cast largely as a kinematical process, and that the self-gravity is
a secondary aspect; it would be interesting to further understand how
these two views relate. But of more immediate practical importance is
to apply the broader insights about dwarf spheroidal stellar
equilibria to observational data, something that we will attempt in
the near future.

\section*{Acknowledgments}
All simulation analysis made use of the pynbody suite
\citep{2013ascl.soft05002P}. AP acknowledges helpful conversations
with James Binney, Simon White, Chervin Laporte and Chiara Tonini and
support from the Royal Society and, during 2013 when a significant
part of this work was undertaken, the Oxford Martin School. JIR
acknowledges support from SNF grant PP00P2\_128540/1. FG acknowledges
support from HST GO-1125, NSF AST-0908499.  NR is supported by STFC
and the European Research Council under the European Community's
Seventh Framework Programme (FP7/2007-2013) / ERC grant agreement no
306478-CosmicDawn.  JD's research is supported by Adrian Beecroft and
the Oxford Martin School. This research used the DiRAC Facility,
jointly funded by STFC and the Large Facilities Capital Fund of BIS.

\bibliographystyle{mn2e}
\bibliography{../bibtex/refs.bib}

\appendix

\section{Background}\label{sec:background}

\begin{figure*}
\begin{center}
\includegraphics[width=0.9\textwidth]{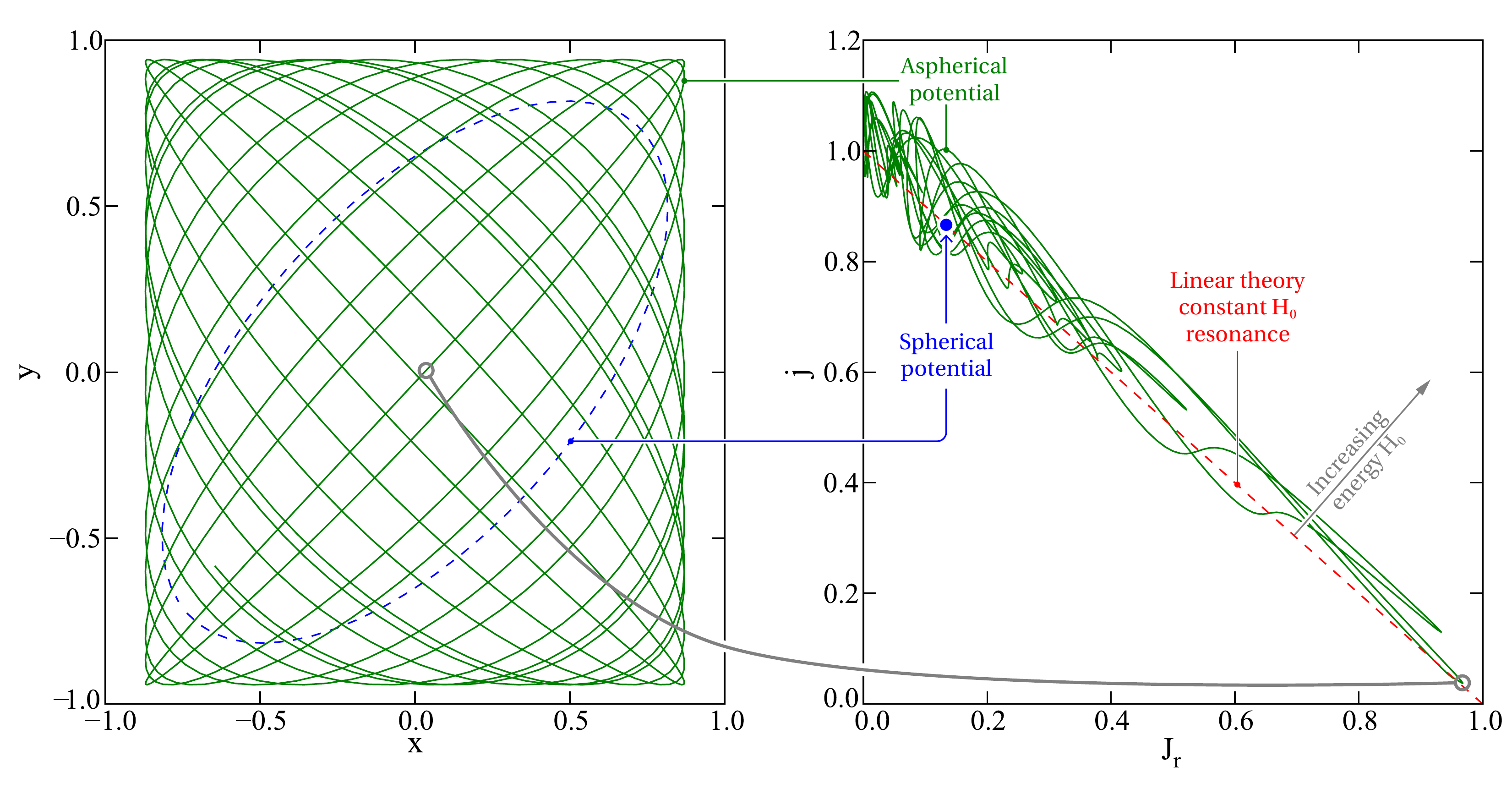}
\end{center}
\caption{An example orbit in a 3D anisotropic harmonic
  oscillator. (Left) the solid line shows a portion of the orbit
  projected into the $x$-$y$ plane. The equivalent orbit in an
  isotropic potential is closed and given by the blue dashed
  line. (Right) the orbit analysed in terms of $J_r$ and $j$, the actions for the
  spherical background Hamiltonian. Because $\Delta H$ is bounded (see
  text) the particle has to stay within a narrow band in the $J_r$-$j$
  plane, corresponding to the linear theory resonance (dotted
  line). The blue dot shows the action for the orbit in the spherical
  potential (for which the actions are by definition fixed).}\label{fig:orbit-Jr-j}
\end{figure*}

\subsection{A brief review of actions}

This Appendix contains a very brief review of actions which are necessary for deriving the main result in the paper. For more complete introductions see \cite{BinneyTremaine2008} or \cite{goldstein2002classical}.
We consider any mechanical
problem described by phase-space coordinates $\vec{q}$ and momenta
$\vec{p}$, with Hamiltonian  $H(\vec{p},\vec{q})$ so that the
equations of motion are
\begin{equation}
\dot{\vec{p}} = -\frac{\partial H}{\partial \vec{q}}\textrm{,}
\hspace{1cm} \dot{\vec{q}} = \frac{\partial H}{\partial
  \vec{p}}\textrm{,}\label{eq:eom}
\end{equation}
where $\dot{\vec{q}}=\dd \vec{q}/\dd t$ and $t$ denotes time.

The actions can be defined starting from any coordinate system in which
the motion is separately periodic in each dimension ({\it i.e.} for
each $i$, $q_i$ and $p_i$ repeat every $\Delta t_i$). The actions $J_i$ are
then given by
\begin{equation}
J_i \equiv \frac{1}{2\pi}\int_0^{\Delta t_i} p_i \dot{q}_i \,\dd t\textrm{ (no sum
  over $i$)}\textrm{.}\label{eq:define-action}
\end{equation}
By construction the actions $J_i$ do not change under time evolution
and are therefore integrals of motion.

We can complete the set of 6 phase-space coordinates with angles $\vec{\Theta}$ in such a way that equations of motion of the form \eqref{eq:eom} apply. Since $J_i$ is constant, we must then have
\begin{equation}
0 = \dot{J}_i=-\frac{\partial H}{\partial \Theta_i}\textrm{,}\hspace{1cm}
\dot{\Theta}_i = \frac{\partial H}{\partial J_i} \equiv \Omega_i(\vec{J})\textrm{,}\label{eq:action-eom}
\end{equation}
where the first equation establishes that $H$ can have no $\Theta$
dependence, and the second that consequently the $\Theta_i$ each
increase at a constant rate in time specified by the frequency
$\Omega_i(\vec{J})$.
The convenience of this set of coordinates is that all the time
evolution of a particle trajectory is represented in a very simple way:
\begin{equation}
J_i = \textrm{constant,} \hspace{1cm} \Theta_i = \textrm{constant} +
\Omega_i\, t\textrm{.}\label{eq:action-eom-integrated}
\end{equation}
Furthermore the equations of motion are canonical, which is sufficient
to demonstrate that the coordinates are canonical -- in other words,
the measure appearing in phase space integrals is $\dd^3 J\,
\dd^3 \Theta$.

We now specialise
to the spherical case, with polar coordinates
$\vec{q}=(r,\theta,\phi)$ and momenta $\vec{p}=(\dot{r},r^2
\dot{\theta}, r^2 \sin^2 \theta \dot{\phi})$. Consider first
$J_{\phi}$,
\begin{equation}
J_{\phi} = \frac{1}{2\pi} \int_0^{\Delta t_{\phi}} \dd t \dot{\phi}^2
r^2 \sin^2 \theta =
\frac{1}{2\pi} \int_0^{2\pi} \dd \phi \dot{\phi} r^2 \sin^2 \theta = j_z
\end{equation}
where $j_z$ is the $z$ component of the specific angular momentum,
$j_z=\dot{\phi} r^2 \sin^2 \theta$ which is a constant
of motion.
Next consider $J_{\theta}$,
\begin{equation}
J_{\theta} = \frac{1}{2\pi} \int_0^{\Delta t_{\theta}} \dd t
\dot{\theta}^2 r^2 = \frac{1}{\pi} \int_{\theta_a}^{\theta_b} \dd \theta
\sqrt{j^2 - \frac{j_z^2}{\sin^2 \theta}}
\textrm{,}
\end{equation}
where $j^2=j_x^2+j_y^2+j_z^2$ is the square of the total specific
angular momentum, and $\theta$ varies between $\theta_a$ and
$\theta_b$ over the course of an orbit. To evaluate the integral
requires a change of variables to relate $\theta_a$ and $\theta_b$ to
the inclination of the orbit and so to $j$ and $j_z$ \citep[e.g.][eq
10.135]{goldstein2002classical}; the final result is that
\begin{equation}
J_{\theta} = j-j_z\textrm{.}
\end{equation}
Because linear combinations of actions are still actions ({\it i.e.}
they still satisfy equation \eqref{eq:action-eom}) one
can take $j_z$ as the first and $j$ as the second action in place of
$J_{\phi}$ and $J_{\theta}$.

The last action is the radial action,
\begin{equation}
J_r = \frac{1}{2 \pi} \int_0^{\Delta t_r} \dd t \dot{r}^2 = \frac{1}{
  \pi} \int_{r_a}^{r_b} \dd r \sqrt{E - j^2/2 r^2 - \Phi(r)}\textrm{,}\label{eq:define-Jr}
\end{equation}
where $\dot{r}$ has been evaluated by energy conservation, and $r$
librates between $r_a$ and $r_b$ over the period of an orbit.


Although the complexity of expression \eqref{eq:define-Jr} appears to
make using the actions cumbersome, the great simplification it brings
to the equations of motion in the background (i.e. equation
\ref{eq:action-eom}) makes the perturbation theory tractable. For that
reason we have used the action-angle coordinates in our analytic
derivation, Section \ref{sec:most-stable-system}, but avoided them
when discussing results from simulations in Section~\ref{sec:ROI-connection}.

\subsection{Perturbed trajectories in the harmonic oscillator case}\label{sec:harmonic-intuition}

Section \ref{sec:most-stable-system} used Hamiltonian perturbation
theory to discuss the behaviour of particles in aspherical
potentials. To connect this more firmly with the equations of motion,
it can be helpful to study orbits in a specific potential and connect
the solutions with the more general statements made by the
perturbation theory. In this Appendix, we use the harmonic oscillator
as such an illustrative example.

Consider the Hamiltonian for a single particle in an anisotropic
harmonic oscillator potential,
\begin{equation}
H = \frac{1}{2} \left(\pi_x^2 + \pi_y^2 + \pi_z^2\right) +
\frac{\omega_0^2}{2} \left( x^2+(1+\epsilon) y^2+(1-\delta) z^2\right)\textrm{,}\label{eq:anisotropic-oscillator}
\end{equation}
where $\pi_x$, $\pi_y$, $\pi_z$ denote the momentum in the three
cartesian directions and without loss of generality $\epsilon\ge 0$
and $\delta\ge 0$.

We want to compare the motion in this potential to the behaviour in a
sphericalised version with Hamiltonian $H_0$; the change
in the Hamiltonian between the true and sphericalised cases is given by
\begin{equation}
\delta H = H-H_0 = \frac{1}{2} \omega_0^2 \left(\epsilon y^2 - \delta
  z^2 \right)\textrm{.}
\end{equation}
We can immediately read off the first result, which is that the
magnitude of $\delta H$ is bounded. The true solution moves around on
a fixed $H$ surface, meaning the fractional error takes a maximum
value given by
\begin{equation}
\frac{\left| \delta H \right |}{H}< \max(\epsilon,\delta)\textrm{.}\label{eq:approx-H0-conserve}
\end{equation}
This is the equivalent of the general statement that $H_0$ variations
are small, given by equation \eqref{eq:H0-variation-small}.

On the other hand, the angular momentum does change significantly. We
can see this as follows: each of $x$, $y$ and $z$ undergoes
oscillation at the frequencies $\omega_0$,
$(1+\epsilon)^{1/2}\omega_0$ and $(1-\delta)^{1/2} \omega_0$
respectively.  Assuming these new frequencies are not commensurate,
the relative phases between the different oscillations slowly shifts
until at some point all three separated oscillators reach $x=0$, $y=0$
and $z=0$ at the same moment. At this point, since the velocity
remains finite, the angular momentum has become zero. It may take a
number of dynamical times before this happens, but (for example) at
the centre of dark matter halos the dynamical time is very short
compared to the Hubble time so angular momentum conservation is
effectively destroyed.

All the above is illustrated in Figure \ref{fig:orbit-Jr-j}.  The left
panel contrasts the orbits for a spherical harmonic oscillator (dashed
line) and an aspherical oscillator (solid line) projected in the
$(x,y)$ plane. The latter obeys equation
\eqref{eq:anisotropic-oscillator} with $\epsilon=\delta=0.1$ (and the
former with $\epsilon=\delta=0$).  The orbits for the spherical case
are closed because the frequencies are identical, so the relative
phase of the $x$ and $y$ part of the motion remains fixed.  Orbits in
the aspherical potential are more complex as the relative phase of the
cartesian components gradually changes; in fact, a particle will
sometimes plunge through the centre of the potential. This is known as
a `box orbit'. In more general triaxial potentials, a variety of orbit
types are possible \citep[e.g.][]{1996ApJ...471...82M}; the importance
of these will be considered momentarily.

The same portion of the orbit is illustrated in the right panel, but
now projected into the spherical actions plane $(J_r,j)$. For the
spherical case, $J_r$ and $j$ are exactly conserved by construction
and the orbit appears as a single point. In the aspherical case (solid
line), $J_r$ and $j$ are not even approximately conserved over more
than a dynamical time. However, even then, the orbit remains close to
the dashed line of constant $H_0$, as required by equation
\eqref{eq:approx-H0-conserve} and more generally by equation
\eqref{eq:H0-variation-small}. From this figure, it is intuitively
plausible that the particle is equally likely to be found anywhere
along the constant $H_0$ contour, which is the essential result of
Section \ref{sec:population}.

Another way to look at the effect is to consider the relationship
between the angular momentum and the harmonic oscillator's cartesian
actions, $J_x$, $J_y$ and $J_z$ which remain constant even in the aspherical
case. Specialising for simplicity to the case that $J_z=0$, one can
show that
\begin{equation}
j = 2 \sqrt{J_x J_y} \sin(\Theta_x-\Theta_y)\textrm{,}
\end{equation}
where $\Theta_x$ and $\Theta_y$ are the angles conjugate to $J_x$ and
$J_y$. Only if $\Theta_x-\Theta_y$ remains constant is $j$ an integral of
motion; as soon as the oscillator is aspherical, one has
\begin{equation}
j = 2 \sqrt{J_x J_y} \sin( \phi_0 + (\Omega_x - \Omega_y ) t)\label{eq:j-not-conserved}
\end{equation}
so that $j$ oscillates on the timescale $2\pi (\Omega_x-\Omega_y)^{-1} \simeq \pi \epsilon^{-1}$.  The situation in 3D is qualitatively similar.

The harmonic oscillator orbits discussed above are all box orbits. More general triaxial potentials support
other orbit types (loop orbits, for example, which are much more
tightly constrained; or chaotic orbits, which are even less tightly
constrained than box orbits). However the Hamiltonian analysis in the main text
is general for all these types of possible orbit.  The fraction of
different orbit types will determine how fast and how far particles
diffuse along the contour.
In realistic cosmological dark matter halos, most orbits in the
central regions are indeed of the box or chaotic type
\citep{2012MNRAS.423.1955Z}. Even with the baryonic
contribution to the gravitational force included (which partially
sphericalises the central potential), the large majority of particles
remain on the same class of orbit as the dark
progenitor \citep{2010MNRAS.403..525V} so long as feedback is strong
enough to prevent long-lived central baryon concentrations developing
\citep{2012MNRAS.422.1863B}.

\subsection{The action space of dark matter halos}\label{sec:Jr-j-halos}

\begin{figure}
\includegraphics[width=0.49\textwidth]{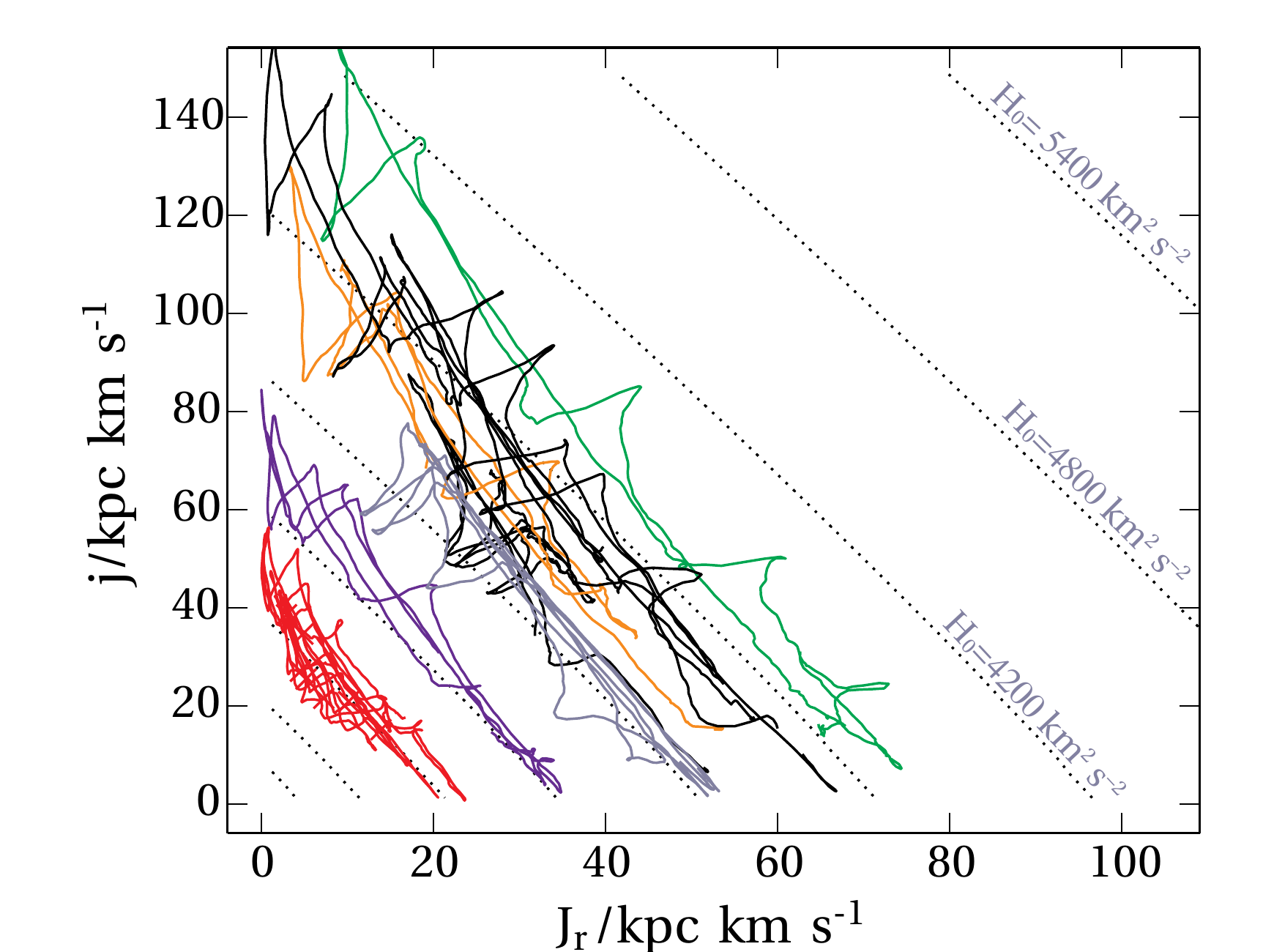}
\caption{Orbits in the action space of the equilibrium `Dwarf' cosmological halo.  Dotted contours of contant spherical specific energy $H_0$ are spaced at $600\,\km^2\,\s^{-2}$. (For this halo $\vmax^2 \simeq 3\,100\,\km^2\,\s^{-2}$.) The straightness of these contours is part of the reason that orbits can efficiently explore the space, as described in the text. }\label{fig:scribble}
\end{figure}

The action-angle space of our equilibrium `Dwarf' dark-matter-only
halo is illustrated in Figure \ref{fig:scribble}, along with some
particle orbits (solid lines) over $1.5\,\Gyr\simeq 3\,\tdyn$. The
Hamiltonian is a function of $J_r$ and $j$ only for any spherical
potential, and so we have suppressed the third action $j_z$.  As
expected, the particles explore the space, approximately running along
the $H_0$ contours, giving rise to the continuous radial orbit
instability described in Section \ref{sec:cont-roi}.  The contours of
$H_0$ give a great deal of dynamical information, because the
background frequencies are defined by $\vec{\Omega}_0 \equiv \partial
H_0 / \partial \vec{J}$. These frequencies are at the core of
perturbation theory through equation \eqref{eq:J-first-order}, but
they also determine the higher-order behaviour as follows.

The first step to understanding the behaviour of resonances is to move
to secular perturbation theory (e.g. section 2.4 of
\citeauthor{1992rsm..book.....L} 1992). Secular analysis splits
resonant orbits into two classes, known as accidentally- and
intrinsically-degenerate. The intrinsic case refers to the situation
where the resonance condition applies globally; in other words, that
$\vec{n} \cdot \vec{\Omega}_0$ is near-constant along lines of
constant $H_0$. Suppose, conversely, that $\vec{\Omega}_0$ did vary
along these directions; then, since $\vec{\Omega}_0$ is defined by the
normal to the $H_0$ contours ($\vec{\Omega}_0 = \partial H_0/\partial
\vec{J}$), this would imply a significant curvature of the dotted
lines in Figure \ref{fig:scribble}. Since there is no such curvature,
we can read off that the frequencies do not change and the dynamics is
in the intrinsically-degenerate regime, giving rise to large-scale
migrations. (The same property also means that the approximation
$\Omega_r = \Omega_r(H_0)$ used in reaching equation
\eqref{eq:mean-j-approx} will be extremely accurate.)  We have
established using numerical investigations that this intrinsic
degeneracy property is generic to any spherical action space with
smooth potentials, rather than being specific to cosmological halos.

\section{The evolution of $f_0$}\label{sec:f0-evolution}

In the main text (Section \ref{sec:most-stable-system}), we showed that a distribution function $f$ is maximally stable to linear perturbations if its sphericalised part $f_0$ appears ergodic, $f_0 = f_0(H_0)$. However we did not discuss the actual evolution of $f_0$ to see whether this limit is likely to be achieved.

This requires time-dependent perturbation theory. We  start by writing the collisionless Boltzmann equation,
\begin{equation}
\frac{\partial f}{\partial t} = [H, f] = [H_0, f_0] + [H_0, \delta f] + [\delta H, f_0] + [\delta H, \delta f]\textrm{,}\label{eq:tdep-boltzmann}
\end{equation}
which is an exact expression. The first term vanishes identically; the second and third terms are linear order, and the final term is second order.
The evolution of $f_0$ is given by taking the time derivative of equation \eqref{eq:f0_from_spherical_average} and interchanging the derivative and integral:
\begin{align}
\frac{\partial f_0}{\partial t} &=
\frac{1}{(2\pi)^3}\int \dd^3 \, \Theta \, [H,f] \nonumber \\
& = \frac{1}{(2\pi)^3}\sum_{\vec{m},\vec{n}} \int \dd^3 \Theta \, \left[\delta H_{\vec{m}}\,e^{i \vec{m} \cdot \vec{\Theta}} ,\, \delta f_{\vec{n}} \,e^{i \vec{n} \cdot \vec{\Theta}} \right]\textrm{,}
\end{align}
where the two linear-order terms have vanished after integrating over $\Theta$. Expanding the Poisson bracket and integrating the remaining term gives
\begin{equation}
\frac{\partial f_0}{\partial t} = - i \sum_{\vec{n}} \vec{n} \cdot \frac{\partial}{\partial \vec{J}} \left( \delta f_{\vec{n}} \delta H_{-\vec{n}} \right)\textrm{.}\label{nonlinear-f0-evolution}
\end{equation}
This shows that the evolution of $f_0$ is a fundamentally non-linear phenomenon. To make further progress, we can eliminate $\delta f_{\vec{n}}$, showing that $f_0$ evolution depends only on $\delta H_{\vec{n}}$. First, Fourier transform the time-dependence of $f_0$ and $\delta f$, so that
\begin{align}
f_0(\vec{J},t) & = \int \dd \omega\, e^{i \omega t}\,
\tilde{f}_0(\vec{J}, \omega)\hspace{0.4cm} \textrm{ and} \\
\delta f(\vec{J}, \vec{\Theta}, t) & = \sum_{\vec{n}} \int \dd \omega \delta\tilde{f}_{\vec{n}} (\vec{J}, \omega)\,e^{i \omega t + i \,\vec{\Theta} \cdot \vec{n}}\textrm{.}
\end{align}
For simplicity, we will restrict attention to the case where $\delta H$ is constant in time. Then the evolution of $\delta f$ is given by the linear-order part of equation \eqref{eq:tdep-boltzmann}, Fourier transformed to give:
\begin{equation}
\left( \omega + \vec{\Omega}_0 \cdot \vec{n} \right) \delta \tilde{f}_{\vec{n}} =  \left(\frac{\partial \tilde{f}_0}{\partial \vec{J}} \cdot \vec{n}\right) \delta H_{\vec{n}}\textrm{.}\label{eq:linear-order-df-dH}
\end{equation}
This is just the time-dependent, Fourier-transformed version of equation \eqref{eq:aspherical-evolution}. Note that, for $\delta H$ and $\delta f$ to be real, the Fourier coefficients must satisfy
\begin{equation}
\delta \tilde{f}_{\vec{n}}(\omega)^* = \delta \tilde{f}_{-\vec{n}}(-\omega) \textrm{ and } \delta H_{\vec{n}}^* = \delta H_{-\vec{n}}\textrm{.}\label{eq:mode-reality}
\end{equation}
These requirements are consistent with the relation \eqref{eq:linear-order-df-dH}.

We can now substitute the linear-order solution
\eqref{eq:linear-order-df-dH} for $\delta f_{\vec{n}}$ in equation
\eqref{nonlinear-f0-evolution} to get the leading-order evolution
equation for $f_0$:
\begin{equation}
\omega \tilde{f}_0 = -\sum_{\vec{n}} \vec{n} \cdot \frac{\partial}{\partial \vec{J}} \left[  \frac{\delta H_{\vec{n}} \delta H_{\vec{-n}}  \vec{n} \cdot \partial \tilde{f}_0/\partial \vec{J}}{\omega + \vec{\Omega}_0 \cdot \vec{n}} \right]\textrm{.}
\end{equation}
Using the reality condition, equation \eqref{eq:mode-reality}, we can
pair up negative and positive $\vec{n}$ modes, giving an alternative
version of the expression that is more explicitly symmetric:
\begin{equation}
\tilde f_0 = \sum_{\vec{n}} \vec{n} \cdot \frac{\partial}{\partial \vec{J}} \left[ \frac{\left| \delta H_{\vec{n}} \right|^2 \vec n \cdot \partial \tilde{f}_0/\partial \vec{J} }{\left| \vec{\Omega}_0 \cdot \vec{n} \right|^2 - \omega^2} \right]\textrm{,}\label{eq:f0-fourier}
\end{equation}
where we have divided both sides by $\omega$ and so the result is
technically only applicable for $\omega \ne 0$. Provided the evolution
of $f_0$ is smooth, its Fourier transform $\tilde{f}_0$ is continuous,
so this is not a problem in practice.

Equation \eqref{eq:f0-fourier} is enough to give some insight into the
relaxation process. For simplicity, consider a perturbation consisting
of a single, resonant $\vec{n}_{\perp}$-mode $\delta
H_{\vec{n}_{\perp}}$ with $\vec{n}_{\perp} \cdot
\vec{\Omega}_0=0$. Then we can explicitly invert the Fourier transform
to yield
\begin{equation}
\frac{\partial^2 f_0}{\partial t^2} = \vec{n}_{\perp} \cdot \frac{\partial}{\partial \vec{J}} \left[ \left| \delta H_{\vec{n}_{\perp}} \right|^2 \vec n_{\perp} \cdot \partial f_0/\partial \vec{J}  \right]\textrm{.}\label{eq:almost-wave-eqn}
\end{equation}
This is a wave equation in the $\vec{n}_{\perp}$ direction with
varying speed-of-sound proportional to $|\delta
H_{\vec{n}_{\perp}}|$. Any unevenness in the resonant directions will
flow away at a speed proportional to the strength of the asphericity,
showing explicitly that $\tilde{f}_0$ evolves towards the spherically
ergodic limit. We hope to investigate the full behaviour of equation
\eqref{eq:f0-fourier} for a variety of regimes in future work.

\end{document}